\documentclass[aps,pre,showpacs,amsfonts,amsmath, amssymb,twocolumn,floatfix, superscriptaddress]{revtex4-1}
\usepackage[final]{graphicx}
\usepackage{xcolor}
\usepackage[colorlinks=true, linkcolor=blue, citecolor=blue, urlcolor=blue]{hyperref}
\usepackage{verbatim}
\usepackage{csquotes}
\usepackage{braket}
\usepackage{mathtools}
\usepackage{bm}
\usepackage{tabularx}
\usepackage{tabu}
\usepackage[normalem]{ulem}

\newcommand{\vct}[1]{\mathbf{#1}}

\newcommand{\com}[1]{}

\newcommand{\be}{\begin{equation}}
\newcommand{\ee}{\end{equation}}

\allowdisplaybreaks

\definecolor{MAGENTA}{rgb}{1.0,0.0,1.0}

\begin{document}


\title{Response of active Brownian particles to shear flow}

\author{Kiryl Asheichyk}
\email[]{asheichyk@is.mpg.de}
\affiliation{4th Institute for Theoretical Physics, Universit\"at Stuttgart, Pfaffenwaldring 57, 70569 Stuttgart, Germany}
\affiliation{Max Planck Institute for Intelligent Systems, Heisenbergstrasse 3, 70569 Stuttgart, Germany}
\author{Alexandre P. Solon}
\email[]{solon@lptmc.jussieu.fr}
\affiliation{Sorbonne Universit\'e, CNRS, Laboratoire de Physique Th\'eorique de la Mati\'ere Condens\'ee, LPTMC, F-75005 Paris, France} 
\author{Christian M. Rohwer}
\email[]{crohwer@is.mpg.de}
\affiliation{4th Institute for Theoretical Physics, Universit\"at Stuttgart, Pfaffenwaldring 57, 70569 Stuttgart, Germany}
\affiliation{Max Planck Institute for Intelligent Systems, Heisenbergstrasse 3, 70569 Stuttgart, Germany}  
\author{Matthias Kr\"uger}
\email[]{matthias.kruger@uni-goettingen.de}
\affiliation{Institute for Theoretical Physics, Georg-August-Universit\"at G\"ottingen, 37073 G\"ottingen, Germany}

\begin{abstract} 
  We study the linear response of interacting active Brownian
  particles in an external potential to simple shear flow.  Using a
  path integral approach, we derive the linear response of any state
  observable to initiating shear in terms of correlation functions
  evaluated in the unperturbed system.  For systems and observables
  which are symmetric under exchange of the $x$ and $y$ coordinates,
  the response formula can be drastically simplified to a form
  containing only state variables in the corresponding correlation
  functions (compared to the generic formula containing also time
  derivatives). In
  general, the shear couples to the particles by translational as well
  as rotational advection, but in the aforementioned case of $xy$
  symmetry only translational advection is relevant in the linear
  regime. We apply the response formulas analytically in solvable cases
  and numerically in a specific setup. In particular, we investigate
  the effect of a shear flow on the morphology and the stress of $N$ confined active
  particles in interaction, where we find that the activity as well as additional alignment interactions generally increase the response.
\end{abstract}


\bibliographystyle{plain}

\maketitle


\section{Introduction}
\label{sec:Introduction}

Many systems found in nature are inherently open and thus operate out
of equilibrium. Among them, active systems are driven by energy
dissipation at the level of each of their individual components. This
strong form of driving at the small scale leads to original collective
behaviors at larger scales, from
flocking~\cite{vicsek1995, Ballerini2008} to
spontaneous bacterial
flows~\cite{Dombrowski2004, Sokolov2007} and many more
(see, e.g., Refs.~\cite{Marchetti2013,Zottl2016,Bechinger2016} for recent reviews).

To better comprehend and manipulate active matter, it is important to
understand how it responds to external perturbations. The response of
a system to shear is of particular interest since it tells us about
the rheology of the fluid under consideration~\cite{Dhont1996, Larson1999}. In that respect, active fluids exhibit surprising
properties. In particular, the
dipolar forces exerted by swimmers in a solution can lead to an
increase~\cite{Rafai2010} or a decrease in
viscosity~\cite{Sokolov2009}, even turning the solution into a
superfluid~\cite{Lopez2015}. This may be of particular
relevance for blood flow \cite{Horner2018}.

The rheology of active fluids can be understood qualitatively through
analytic arguments~\cite{Hatwalne2004,Takatori2017} and
numerical simulations~\cite{Saintillan2010,Fielding2011} but the
quantitative prediction of transport coefficients remains mostly out
of reach. To this aim, response relations are particularly useful
since they allow to compute the average response of a system in terms
of correlation functions in the unperturbed system. When considering
perturbations of an equilibrium system, these are the celebrated
fluctuation-dissipation theorems~\cite{Callen1951, Weber1956, Kubo1966} and Green-Kubo relations \cite{Green1954, Kubo1966,
  Kubo1991}. Extending such relations to describe the perturbation of
nonequilibrium steady states has been the subject of intense
research~\cite{Harada2005, Speck2006, Blickle2007, Chetrite2008,
  Baiesi2009_1, Baiesi2009_2, Baiesi2010, Prost2009, Kruger2009, Seifert2010, Seifert2012,
  Warren2012, Caprini2018} and is the topic of the present article. Note that
activity itself is sometimes treated as a perturbation
parameter~\cite{Fodor2016, Sharma2016, Merlitz2018}. This is not the case here as
we consider perturbations close to nonequilibrium steady states with
potentially high activity.

In this work, we study the linear response to shear of a collection of
active, i.e., self-propelled, Brownian particles (ABPs). This model of ``dry'' active matter (neglecting the
effect of a solvent) has become a workhorse for studying active systems~\cite{Zottl2016, Bechinger2016}, 
including important phenomena such as motility induced phase separation~\cite{Fily2012} which was also found for colloidal
swimmers~\cite{Buttinoni2013}. The microscopic mechanisms of swimming motion, and the effects of hydrodynamic interactions are yet a diverse field of research, see, e.g., 
Refs.~\cite{Marchetti2013, Zottl2016, Zottl2012, Redding2013, Hu2015, Matas-Navarro2014, Furukawa2014, Uspal2015}.

Using path integral techniques, we derive here a
general formula, valid for any interaction and external potentials, to
compute the linear response to shear in terms of correlation functions
evaluated in the unsheared system. In the special case where the
situation is symmetric under exchange of the $x$ and $y$ axes, the
response formula simplifies such that it contains
only static variables (no velocities) and is not affected by shear rotation. As we
show, the formula recovers results previously derived in the
literature for one ABP in a shear flow in free
space~\cite{tenHagen2011} or confined by a harmonic
potential~\cite{Li2017}. Our work complements that of Ref.~\cite{Szamel2017}, which derived, in another model of active particles,
a response relation for perturbations via a potential (not shear). Further, we apply our response formula to $N$ interacting
particles confined in a harmonic potential in two space dimensions. We compute the average of
$ \sum_{i=1}^Nx_iy_i $, where $(x_i,y_i)$ is the position of particle
$i$, to quantify the effect of shear on the morphology of the
suspension. We also numerically check the consistency of our results,
and therefore the response formula, by computing the response directly
in the sheared system.

The paper is organized as follows. In Sec.~\ref{sec:System_and_Model},
we describe the model and introduce relevant physical
quantities. Section~\ref{sec:Linear_Response_to_Shear_Flow} contains a
detailed derivation and discussion of the response formulas. The
formulas are then applied analytically to single-particle systems in
Sec.~\ref{sec:Analytical_Examples}, and numerically to a many-body
situation in Sec.~\ref{sec:Numerical_Example}. We close this
manuscript by a summary and conclusions in Sec.~\ref{sec:Conclusion}.

\section{System and Model}
\label{sec:System_and_Model}
\begin{figure}[!t]
\includegraphics[width=0.8\linewidth]{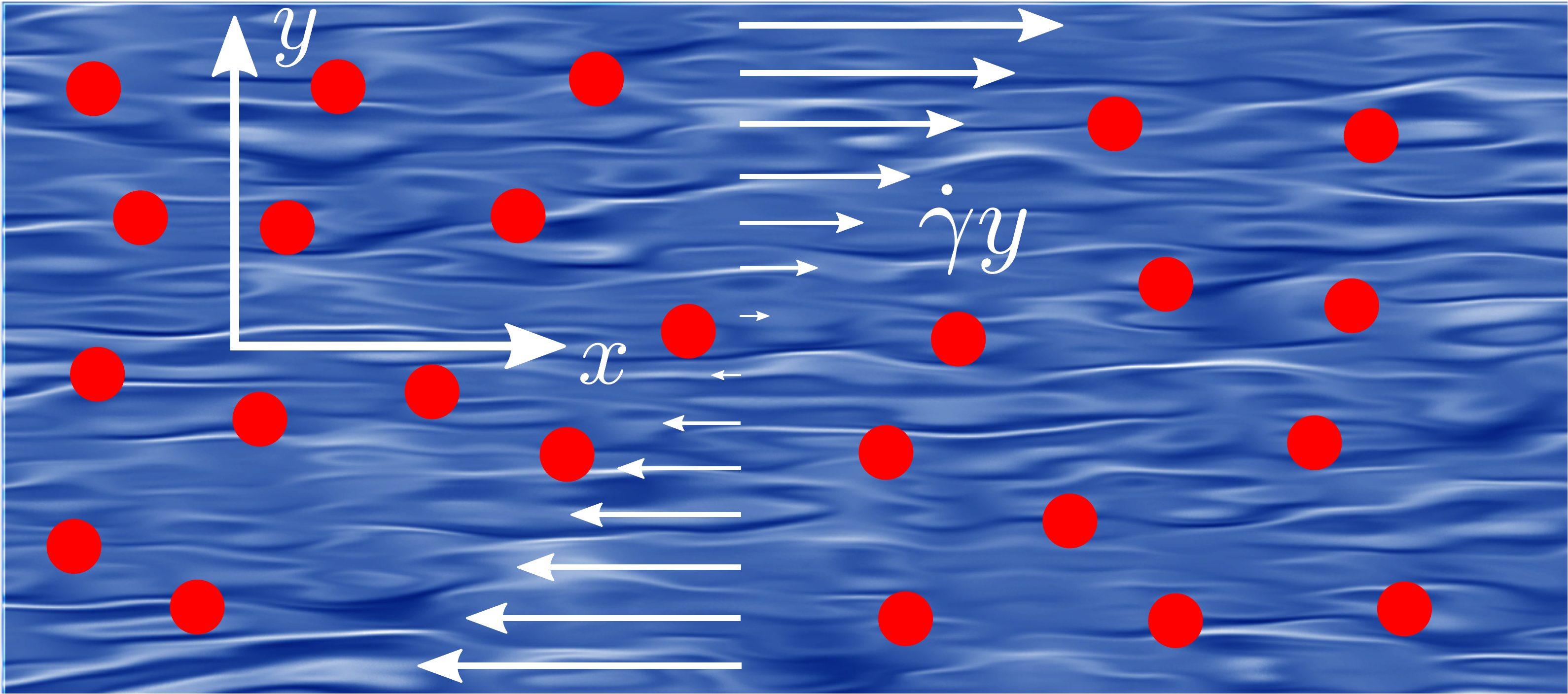}\\
\vspace{0.3cm}
\includegraphics[width=0.8\linewidth]{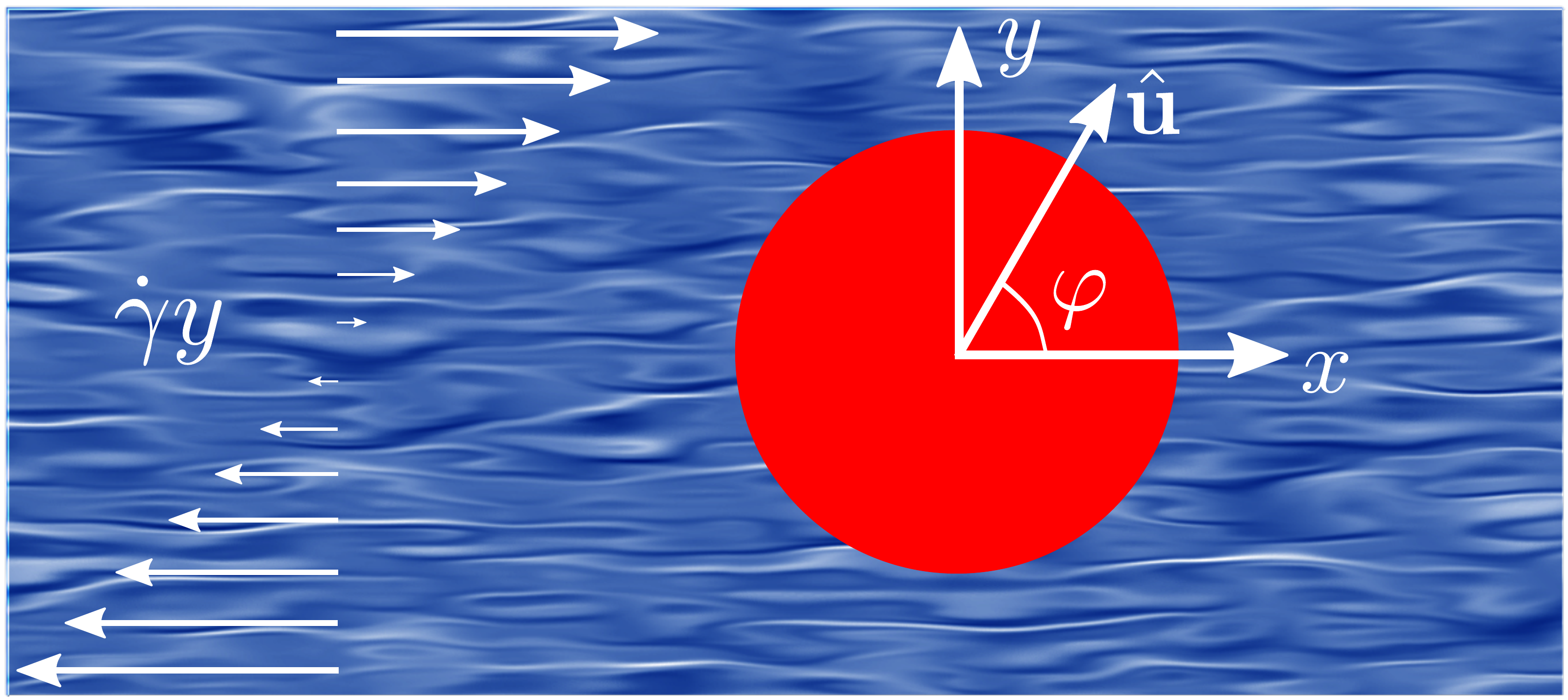}
\caption{\label{fig:System} The system under consideration:
  interacting active Brownian particles (red circles) exposed to
  simple shear flow (white arrows) with shear velocity
  $ \dot{\gamma}y $. {\bf Top:} Schematic representation of $ N $
  particles. {\bf Bottom:} Schematic representation of a single
  particle with heading $ \hat{\vct{u}} $.}
\end{figure}

We consider $ N $ overdamped active Brownian particles, subject to
interactions and external forces. For simplicity, we provide the derivation and explicit examples for a two-dimensional system, which is relevant to many
experiments~\cite{Deseigne2010, Ginot2015, dAlessandro2017,
  Hennes2017}. The generalization to ABPs in 3D and also to more general active particle models is given in Subsec.~\ref{subsec:General}. Each particle thus has three degrees of freedom: two
translational (in the $ x $ and $ y $ directions) and an angle
$ \varphi $ parametrizing its heading
$ \hat{\vct{u}}(\varphi)=\left(\cos(\varphi), \sin(\varphi)\right)^T $
in which the particle self-propels with velocity
$ v_0\hat{\vct{u}}(\varphi) $. The particles are further exposed to a
simple shear flow pointing in the $ x $ direction, with shear velocity
$ \dot{\gamma}y $, where $ \dot{\gamma} $ is shear rate. A schematic
representation of the system is given in Fig.~\ref{fig:System}.

The flow both advects the particles and rotates them~\cite{Dhont1996}.
The dynamics of the $ i $th particle is then given by the overdamped
Langevin equations~\cite{tenHagen2011, Li2017}:
\begin{subequations}\label{eq:LEs}
\begin{align}
&\dot{x}_i = \dot{\gamma}y_i + v_0\cos\varphi_i + \mu_{\rm t} F_{ix}^{\rm int} + \mu_{\rm t} F_{ix}^{\rm ext} + \mu_{\rm t} f_{ix} \label{eq:LE1},\\
&\dot{y}_i = v_0\sin\varphi_i + \mu_{\rm t} F_{iy}^{\rm int} + \mu_{\rm t} F_{iy}^{\rm ext} + \mu_{\rm t} f_{iy} \label{eq:LE2},\\
&\dot{\varphi}_i= -\frac{\dot{\gamma}}{2} + \mu_{\rm_r}(M_i + g_i) \label{eq:LE3},
\end{align}
\end{subequations}
where $ F_{i\alpha}^{\rm int}$ and $ F_{i\alpha}^{\rm ext} $ denote
inter-particle and external forces on particle $i$ in direction
$\alpha$. We allow for very general forces that need not arise from
potentials and can depend on positions and orientations of the
particles, i.e., 
$ F_{i\alpha}^{\rm int} = F_{i\alpha}^{\rm int}(x_1, \dots, x_N, y_1,
\dots, y_N, \varphi_1, \dots, \varphi_N) $ and
$ F_{i\alpha}^{\rm ext} = F_{i\alpha}^{\rm ext}(x_i, y_i,
\varphi_i)$. $ M_i =M_i(x_1, \dots, x_N, y_1,
\dots, y_N, \varphi_1, \dots, \varphi_N)$ is the torque acting on particle $ i $, which can also depend on positions and orientations of all particles. The first term on the right-hand side of
Eq.~\eqref{eq:LE3} is the aforementioned rotation due to shear, where
the prefactor of $\frac{1}{2}$ can be derived by considering an
isolated particle in shear flow~\cite{Dhont1996}. $ \mu_{\rm t} $ and $ \mu_{\rm r} $ denote microscopic translational and rotational mobilities, respectively. The stochastic terms
$ f_{i\alpha} $ and $ g_i $ are uncorrelated Gaussian white noises
with moments
\begin{subequations}\label{noise_properties}
\begin{align}
\!\! \!\langle f_{i\alpha}(t) \rangle& = 0,\!\!\!&\langle f_{i\alpha}(t)f_{j\beta}(t') \rangle &= \frac{2D_{\rm t}}{\mu_{\rm t}^2} \delta_{ij}\delta_{\alpha \beta}\delta(t-t'), \label{eq:random_force_properties} \\
\langle g_i(t) \rangle& = 0,  &\langle g_i(t)g_j(t') \rangle &= \frac{2D_{\rm r}}{\mu_{\rm r}^2}\delta_{ij}\delta(t-t'), \label{eq:random_torque_properties}
\end{align}
\end{subequations}
where $ D_{\rm t} $ and $ D_{\rm r} $ are the translational and
rotational diffusion coefficients and the averaging
$ \langle \dots \rangle $ is with respect to noise realizations.

\section{Linear Response to Shear Flow}
\label{sec:Linear_Response_to_Shear_Flow}
\subsection{Preliminaries}
\label{subsec:Preliminaries}
We want to compute the linear response of the collection of active
Brownian particles described in the previous section to shear flow,
applied for $ t \geq 0 $. We denote the state of the system at time
$ t $ by
$\Gamma(t)\equiv \{x_1(t), \dots, x_N(t), y_1(t), \dots, y_N(t),
\varphi_1(t), \dots, \varphi_N(t)\}$ and introduce the following
averages of a state observable $ A (t) \equiv A(\Gamma(t)) $:
$ {\langle A(t) \rangle}_0 $, $ {\langle A \rangle}_{\rm st} $,
$ {\langle A(t) \rangle}^{(\dot{\gamma})}_0 $,
$ {\langle A(t) \rangle}_{\rm st}^{(\dot{\gamma})} $.  The lower
index, $\langle \dots \rangle_0$ or $\langle \dots \rangle_{\rm st}$,
indicates that the system either started in a specific configuration
$ \Gamma(0) $ at $t=0$ or was in the unperturbed steady (stationary) state at
$t=0$, respectively. In the latter case, the averaging involves averaging over the initial steady-state ensemble. The upper index
$\langle \dots \rangle^{(\dot\gamma)}$ indicates that the system is
sheared for $t \geq 0 $, while $\langle \dots \rangle$ denotes an
average in the unsheared system. This notation is used in the same way
for time-dependent correlations $ C(t, t') \equiv A(t)B(t')$. See
Table~\ref{table:averages} for details.

\begin{table}[!t]
\caption{\label{table:averages}Description of the different types of averaging}
\begin{ruledtabular}
\begin{tabularx}{0.3\textwidth}{l|X}
  Average ($ t \geq t' \geq 0 $) & Description\\
  \hline\\
  $ {\langle A(t) \rangle}_0, {\langle C(t, t') \rangle}_0 $ & Average over noise realizations in the unsheared system given a specific initial condition $ \Gamma(0) $ at time $ t = 0 $ \\
  \hline\\
  $ {\langle A \rangle}_{\rm st}, {\langle C(t-t') \rangle}_{\rm st} $ & Average over noise realizations in steady state of the unsheared system\\
  \hline\\
  $ {\langle A(t) \rangle}^{(\dot{\gamma})}_0 $ & Average over noise realizations in the sheared system given a specific initial condition $ \Gamma(0) $ at time $ t = 0 $ \\
  \hline\\
  $ {\langle A(t) \rangle}_{\rm st}^{(\dot{\gamma})} $ & Average over noise realizations in the sheared system given a steady-state ensemble at time $ t = 0 $. Letting $ t \to \infty $, $ {\langle A(t) \rangle}_{\rm st}^{(\dot{\gamma})} $ then approaches its steady state value under shear\\
\end{tabularx}
\end{ruledtabular}
\end{table}

In the following, we compute the responses
$ {\langle A(t) \rangle}_0^{(\dot{\gamma})} - {\langle A(t) \rangle}_0
$ and
$ {\langle A(t) \rangle}_{\rm st}^{(\dot{\gamma})} - {\langle A
  \rangle}_{\rm st} $ to linear order in $ \dot{\gamma} $ using the
path integral representation of the dynamics.

\subsection{Linear response from the path integral representation}
\label{subsec:LR_general}
Linear response theory using path integrals has been treated previously, see Ref.~\cite{Baiesi2009_2}. We provide the following derivation for completeness. For conciseness, we use more general notations in this section. We
consider a system of $ \widetilde{N} $ stochastic variables
$ \Gamma = \{x_1, \dots, x_{\widetilde{N}}\}$ obeying coupled Langevin
equations,
\begin{equation}
\dot{x}_i = F_i(\Gamma(t), t, \lambda) + f_i(t), \quad i = 1, \dots , \widetilde{N},
\label{eq:LEs_general}
\end{equation}
where $F_i(\Gamma(t), t, \lambda)$ can depend explicitly on time as
well as an external parameter $ \lambda $. The noises $ f_i $ are
independent Gaussian white noises with moments
\begin{equation}
\langle f_i(t) \rangle = 0, \quad \langle f_i(t)f_j(t') \rangle = 2\alpha_i\delta_{ij}\delta(t-t').
\label{eq:random_noise_properties_general}
\end{equation} 
For simplicity, we assume the noise variance $ \alpha_i $ to be
independent of $\Gamma$ and $t$, so that we do not need to specify an
interpretation for the stochastic equations. Extension to the case
where $ \alpha_i $ depends on $ \Gamma $ (multiplicative noise) is
straightforward, although more technical~\cite{Cugliandolo2017}.

As before, we denote by $ \Gamma(t) $ the state of the system at time
$ t $ for a given noise realization. In addition, we denote by
$ \{\Gamma\} $ the full history (the path) of the system on the time
interval $ [0,t] $ for a given noise realization. The average
$ {\langle A(\Gamma(t)) \rangle}_0^{(\lambda)} $ given an initial
condition $ \Gamma(0) $ can then be written as~\cite{Altland2010}
\begin{equation}
{\langle A(\Gamma(t)) \rangle}_0^{(\lambda)} = \int_{\Gamma(0)} D\Gamma A(\Gamma(t)) W^{(\lambda)}(\{\Gamma\}),
\label{eq:PI_initial}
\end{equation}
where $ D\Gamma \equiv \prod_{i=1}^{\widetilde{N}}Dx_i $ is the functional
integration measure and $ W^{(\lambda)}(\{\Gamma\}) $ is the path
weight. The latter follows from standard procedures, e.g., the
Martin-Siggia-Rose-Janssen-de Dominicis (MSRJD)
approach~\cite{Martin1973, Janssen1976, DeDominicis1978,Altland2010}
or the Onsager-Machlup approach~\cite{Onsanger1953,
  Machlup1953}. Defining
\begin{equation}
\mathcal{X}_i(t,\lambda) \equiv \dot x_i(t) - F_i(\Gamma(t), t, \lambda),
\label{eq:Xdef}
\end{equation}
one obtains for the path weight the celebrated Onsager-Machlup functional~\cite{Onsanger1953}
\begin{align}
\notag& W(\{\Gamma\})^{(\lambda)} \propto e^{-\mathcal A(t, \lambda)},\\
& \mathcal{A}(t, \lambda) =\int_0^t dt' \sum_{i=1}^{\widetilde{N}}\frac{1}{4\alpha_i}\mathcal{X}^2_i(t',\lambda),
\label{eq:W_initial}
\end{align}
where $ \mathcal{A} $ is the action of the system \footnote{ The
  action is generically quadratic in $\mathcal{X}_i$ due to the
  Gaussian nature of the noise.  }, and an underlying It\=o
discretization has been employed.  We can now expand the path weight
in powers of $ \lambda $,
\begin{equation}
W^{(\lambda)}(\{\Gamma\}) = W(\{\Gamma\})\left[ 1+ \lambda \int_0^tdt'B(t') \right]+\mathcal{O}(\lambda^2),
\label{eq:W_expansion}
\end{equation}
where $ W(\{\Gamma\}) $ is the path weight of the unperturbed system
(given by Eq.~\eqref{eq:W_initial} with $ \lambda = 0 $), and we have
defined
\begin{align}
\notag &B(t) = -\sum_{i=1}^{\widetilde{N}}\frac{1}{2\alpha_i}\left[\mathcal X_i(t,\lambda)\frac{\partial\mathcal X_i(t,\lambda)}{\partial\lambda}\right]\Bigg|_{\lambda=0}\\
&= \sum_{i=1}^{\widetilde{N}}\frac{1}{2\alpha_i}\left\{\left[\dot x_i(t) - F_i(\Gamma(t), t, \lambda)\right]\frac{\partial F_i(\Gamma(t), t, \lambda)}{\partial \lambda}\right\}\Bigg|_{\lambda=0}.
\label{eq:B}
\end{align}
Note that $ \dot x_i(t) - F_i(\Gamma(t), t, \lambda) $ can be replaced by
$ f_i(t) $ according to Eq.~\eqref{eq:LEs_general}.  Inserting
Eq.~\eqref{eq:W_expansion} into Eq.~\eqref{eq:PI_initial}, we find the
linear response formula
\begin{equation}
{\langle A(t) \rangle}_0^{(\lambda)} -{\langle A(t) \rangle}_0 = \lambda\int_0^tdt'{\langle A(t)B(t') \rangle}_0.
\label{eq:LR_general_incon}
\end{equation}
For a system initially in steady state, one has a similar result, but
with averages conditioned on being in steady state at time $ t = 0 $, i.e.,
\begin{equation}
{\langle A(t) \rangle}_{\rm st}^{(\lambda)} -{\langle A \rangle}_{\rm st} = \lambda\int_0^tdt'{\langle A(t)B(t') \rangle}_{\rm st}.
\label{eq:LR_general_instead}
\end{equation}
We emphasize that the unperturbed system does not have to be in
equilibrium for Eqs.~\eqref{eq:LR_general_incon} and
\eqref{eq:LR_general_instead} to hold, so that the above derivation
encompasses the case of active particles perturbed by shear, as
presented in Sec.~\ref{sec:System_and_Model}, where $\lambda$
corresponds to the shear rate $ \dot{\gamma} $ and $ W(\{\Gamma\}) $
is the path weight of the unsheared \textit{active} system.

Eqs.~\eqref{eq:LR_general_incon} and~\eqref{eq:LR_general_instead}
have been derived previously via different means \cite{Baiesi2009_1, Baiesi2009_2,
  Baiesi2010, Warren2012} (see Eqs.~(2) and~(3) in
Ref.~\cite{Warren2012}) and are thus in agreement with previous works.

It is worth noting that for a perturbation via a potential
$ V(\Gamma(t), \lambda) $, one has
$-\mathcal X_i\partial_\lambda \mathcal X_i \sim
-(\dot{x}_i-F_i)\partial_{x_i}\partial_{\lambda} V$, where the first
term, $-\dot{x}_i \partial_{x_i}\partial_\lambda V$, is
time-antisymmetric and the second term,
$ F_i \partial_{x_i}\partial_\lambda V$, is time-symmetric. If,
additionally, the unperturbed steady state obeys detailed balance, the
two terms yield equal contributions \cite{Baiesi2009_1, Baiesi2009_2, Basu2015}, and
$ B $ in Eq.~\eqref{eq:B} can be written as a total time derivative
(using stochastic calculus with care \cite{Gardiner2010,
  Wynants2010}): the fluctuation-dissipation theorem is recovered.

We thus emphasize that using response theory on sheared active systems
is challenging for two reasons: shear cannot be written as a potential
perturbation, and the unperturbed system does not obey detailed
balance.

\subsection{Linear response of active particles to shear}
\label{subsec:LR_our_system}
We now apply the response formulas of the previous subsection to the case
of active Brownian particles under shear, introduced in
Sec.~\ref{sec:System_and_Model}, and look at the response to small
shear rate $\dot{\gamma}$.  Explicitly, we find that
Eq.~\eqref{eq:W_initial}, for the dynamics described by
Eqs.~\eqref{eq:LEs}, gives the Onsager-Machlup action
\begin{widetext}
\begin{align}
\mathcal A(t, \dot{\gamma})=\frac{1}{4}\int_0^tdt'\sum_{i=1}^N\Bigg[&\frac{1}{D_{\rm t}}\Big[\left(\dot{x}_i(t')-\dot{\gamma}y_i(t')-v_0\cos\varphi_i(t')-\mu_{\rm t} F_{ix}^{\rm int}(t')-\mu_{\rm t} F_{ix}^{\rm ext}(t')\right)^2\nonumber\\
&+ \left(\dot{y}_i(t')-v_0\sin\varphi_i(t')-\mu_{\rm t} F_{iy}^{\rm int}(t') - \mu_{\rm t} F_{iy}^{\rm ext}(t')\right)^2\Big]+\frac{1}{D_{\rm r}}\left(\dot{\varphi}_i(t')+\frac{\dot{\gamma}}{2}-\mu_{\rm r}M_i(t')\right)^2\Bigg].
\label{eq:W_shear}
\end{align}
By identifying the function $ B $ according to Eq.~\eqref{eq:B} and inserting it in expressions~\eqref{eq:LR_general_incon} 
and~\eqref{eq:LR_general_instead}, one obtains the linear response to shear flow,
\begin{align}
\notag{\langle A(t) \rangle}_{0({\rm st})}^{(\dot{\gamma})} & - {\langle A(t) \rangle}_{0({\rm st})}\\
& = \frac{\dot{\gamma}}{4}\int_0^tdt'{\left\langle A(t)\sum_{i=1}^N\left\{\frac{2}{D_{\rm t}}\left[\dot{x}_i(t')-v_0\cos\varphi_i(t') - \mu_{\rm t} F_{ix}^{\rm int}(t')-\mu_{\rm t} F_{ix}^{\rm ext}(t')\right]y_i(t') - \frac{1}{D_{\rm r}}\left[\dot{\varphi}_i(t')-\mu_{\rm r}M_i(t')\right]\right\}\right\rangle}_{0({\rm st})},
\label{eq:LR_final}
\end{align}
\end{widetext}
where the subscript \enquote{$ 0({\rm st}) $} indicates the average
given that the unperturbed system is either in a state with an initial
condition $ \Gamma(0) $ or in steady state (see
Table~\ref{table:averages}).

The result of Eq.~\eqref{eq:LR_final} generalizes traditional
Green-Kubo relations~\cite{Green1954, Kubo1966, Kubo1991} to
shear perturbations of active Brownian particles. Indeed, in the
passive limit, Eq.~\eqref{eq:LR_final} reduces to equilibrium relation
(the limit is obtained by setting the terms $ \propto v_0 $ and
$ \propto \frac{1}{D_{\rm r}} $ to zero). The term
$ \propto \frac{1}{D_{\rm t}} $ is the response due to the advection of
the particles by the shear flow (described by the term $ \dot{\gamma}y_i $ in
Eq.~\eqref{eq:LE1}), while the term $\propto \frac{1}{D_{\rm r}} $ is
the response due to the rotation of the particles (described by the term
$ -\frac{\dot{\gamma}}{2} $ in Eq.~\eqref{eq:LE3}). Thus we see that,
at linear order, shear translation and shear rotation do not couple,
as expected.

Eq.~\eqref{eq:LR_final} becomes more intuitive when introducing the
following pseudo-force acting on particle $i$ in Langevin
equations~\eqref{eq:LEs},
\begin{equation}
\widetilde{\vct{F}}_i= \vct{F}_i^{\rm s}  +\vct{F}_i^{\rm int}+ \vct{F}_i^{\rm ext}.
\label{eq:pseudo_force}
\end{equation}
Here, we have formally interpreted the self-propulsion velocity $ v_0\hat{\vct{u}}_i $ as a force, $\vct{F}_i^{\rm s} = \frac{v_0}{\mu_{\rm t}}\hat{\vct{u}}_i$. 
Inspired by the conventional (interaction) stress tensor~\cite{Fuchs2005}
\begin{equation}
\sigma^{\rm int}_{xy}=-\sum_{i=1}^N F_{ix}^{\rm int} y_i,
\label{eq:stress}
\end{equation}
which appears for sheared passive bulk systems, we introduce a
pseudostress tensor, whose $ xy $ component reads
\begin{equation}
\widetilde{\sigma}_{xy}= -\sum_{i=1}^N \widetilde{F}_{ix}y_i = {\sigma}^{\textrm s}_{xy}+{\sigma}^{\rm int}_{xy}+{\sigma}^{\textrm{ext}}_{xy},
\label{eq:pseudo_stress}
\end{equation}
where ${\sigma}^{(\cdots)}_{xy}\equiv -\sum_{i=1}^N F_{ix}^{(\cdots)} y_i$. In accordance with Eq.~\eqref{eq:pseudo_force}, $\widetilde{\sigma}_{xy}$ additionally contains external forces and the self-propulsion (swim) force, as in Ref.~\cite{Takatori2014}.
Eq.~\eqref{eq:LR_final} then acquires the form
\begin{align}
\notag{\langle A(t) \rangle}_{0({\rm st})}^{(\dot{\gamma})} &- {\langle A(t) \rangle}_{0({\rm st})} =\frac{\dot{\gamma}\mu_{\rm t}}{2D_{\rm t}} \int_0^t dt'{\left\langle A(t)\widetilde{\sigma}_{xy}(t')\right\rangle}_{0({\rm st})}\notag\\
\notag&+\frac{\dot{\gamma}}{4}\int_0^tdt'\Bigg\langle A(t)\sum_{i=1}^N\Bigg\{\frac{2}{D_{\rm t}}\dot{x}_i(t')y_i(t')\\
&-\frac{1}{D_{\rm r}}\left[\dot{\varphi}_i(t')-\mu_{\rm r}M_i(t')\right]\Bigg\}\Bigg\rangle_{0({\rm st})}.
\label{eq:LR_final_stress_form}
\end{align}
The terms in Eq.~\eqref{eq:LR_final_stress_form} have clear meanings in terms of the various contributions to the pseudostress tensor $\widetilde{\sigma}_{xy}$ (compare the response formula for passive 
bulk systems in terms of only ${\sigma}^{\rm int}_{xy}$~\cite{Fuchs2005}).
Furthermore, a term
$\sim\dot{x_i }y_i$ appears: it is time-antisymmetric, and is present
because the unsheared system does not obey detailed balance. Last, the
term $\sim \dot{\varphi_i}-\mu_{\rm r}M_i$ appears because shear rotates the
particles, as mentioned before.

Finally, we note that, since we used the It\=o
convention~\cite{Altland2010} to discretize the Langevin equations in
deriving actions~\eqref{eq:W_initial} and~\eqref{eq:W_shear}, the
integrals in the response
formulas~\eqref{eq:LR_general_incon},~\eqref{eq:LR_general_instead},
and~\eqref{eq:LR_final} are stochastic It\=o
integrals~\cite{Gardiner2010, Wynants2010}. However, one can show that, in the case of Eq.~\eqref{eq:LR_final}, 
these are equivalent to stochastic Stratonovich integrals, and can
hence be treated as standard Riemann integrals~\cite{Gardiner2010,
  Wynants2010}.

\subsection{Response formula in terms of state variables using symmetries}
\label{subsec:LR_another_form}
The linear response formula~(\ref{eq:LR_final_stress_form}) contains
instantaneous velocities, $ \dot{x}_i $ and $ \dot{\varphi}_i $, which are not present in the Green-Kubo formula for overdamped passive systems. While
these derivatives emerge from a well-defined procedure and can be
measured in computer simulations, they are typically not measurable in
experiments. We thus aim to give
formula~\eqref{eq:LR_final_stress_form} in terms of state quantities
which we expect to be more easily accessible.  This is possible for a
specific class of systems and observables.

In order to obtain a total time derivative from the term
$\dot x_i y_i$, we add to Eq.~\eqref{eq:LR_final_stress_form} the case
of shear flow in $y$ direction with gradient in $x$, denoting averages
of the latter by
$ {\langle A(t) \rangle}_{0({\rm st})}^{(\dot{\gamma}, y)} $. This
yields $\dot x_i y_i+\dot y_i x_i$, and, because the stochastic It\=o
integrals in Eq.~\eqref{eq:LR_final_stress_form} are equivalent to stochastic
Stratonovich integrals~\cite{Gardiner2010, Wynants2010}, we identify
the total time derivative
\begin{align}
\notag & \int_0^tdt' \big[\dot{x}_i(t')y_i(t')+\dot{y}_i(t')x_i(t')\big]\\
& = \int_0^tdt' \frac{{\rm d} (x(t')y(t'))}{{\rm d}t'} = x_i(t)y_i(t) - x_i(0)y_i(0).
\label{eq:LRAF_xy}
\end{align}
We further note that the two shear flows exert opposite torques, such
that there is no net shear rotation when both are applied. This fact
and Eq.~\eqref{eq:LRAF_xy} allow us to remove time derivatives
from the response formula~\eqref{eq:LR_final_stress_form}. We then
have
\begin{align}
\notag {\langle A(t) \rangle}&_{0({\rm st})}^{(\dot{\gamma})}+{\langle A(t) \rangle}_{0({\rm st})}^{(\dot{\gamma}, y)} - 2{\langle A(t) \rangle}_{0({\rm st})}\\
\notag = & \ \frac{\dot\gamma \mu_{\rm t}}{2 D_{\rm t}} \int_0^t dt'\left\langle A(t)\left[\widetilde{\sigma}_{xy}(t')+\widetilde{\sigma}_{yx}(t')\right]\right\rangle_{0({\rm st})}\\
&+\frac{\dot{\gamma}}{2D_{\rm t}}{\left\langle A(t)\sum_{i=1}^N\left[{x}_i(t)y_i(t)  - {x}_i(0)y_i(0) \right]\right\rangle}_{0({\rm st})}.
\label{eq:LRAF1}
\end{align}
This simplification can be traced to the fact that, when the shear
flows in $x$ and $y$ are superimposed, the perturbation can be seen
exactly as arising from a potential
$ V(\Gamma, \dot{\gamma}) = -\frac{\dot{\gamma}}{\mu_{\rm t}}\sum_{i=1}^Nx_i y_i $.

For certain special cases, we can further simplify Eq.~\eqref{eq:LRAF1}. We therefore consider the case where a steady active system is perturbed by shear for $ t \geq 0 $. We also restrict to systems which are $ xy $ symmetric, i.e., the systems for which the external and interaction potentials, giving rise to $ \vct{F}^{\rm ext} $ and $ \vct{F}^{\rm int} $, respectively, as well as torques $ M_i $ are symmetric under interchange of  $ x $ and $ y $ (imagine interacting particles in a square box). These criteria allow, e.g.,  for aligning interactions between particles, as used in Eq.~\eqref{eq:alignment} below. If, additionally, the observable $ A $ is also symmetric under interchange of $ x $ and $ y $, e.g., $ A = \sum_{i=1}^Nx_iy_i $, then, by symmetry, the responses to shear flow in the $x$ and $y$ directions are equal,
$ {\langle A(t) \rangle}_{\rm st}^{(\dot{\gamma})} = {\langle A(t)
  \rangle}_{\rm st}^{(\dot{\gamma}, y)} $. Eq.~\eqref{eq:LRAF1} then takes the desired form of response to shear,
\begin{align}
\notag {\langle A(t) \rangle}_{\rm st}^{(\dot{\gamma})} & -  {\langle A \rangle}_{\rm st}= \frac{\dot\gamma \mu_{\rm t}}{4 D_{\rm t}} \int_0^t dt'\left\langle A(t)\left[\widetilde{\sigma}_{xy}(t')+\widetilde{\sigma}_{yx}(t')\right]\right\rangle_{\rm st}\\
&+\frac{\dot{\gamma}}{4D_{\rm t}}{\left\langle A(t)\sum_{i=1}^N\left[{x}_i(t)y_i(t)  - {x}_i(0)y_i(0) \right]\right\rangle}_{\rm st}.
\label{eq:LRAF2}
\end{align}
Note that the term
$ {\left\langle A(t)\sum_{i=1}^Nx_i(0)y_i(0)\right\rangle}_{\rm st} $
is a stationary correlation function with time difference $ t $, while
the
term$ {\left\langle A(t)\sum_{i=1}^Nx_i(t)y_i(t)\right\rangle}_{\rm
  st} $ is a stationary equal-time correlation function, and is thus
time independent.

As a final simplification, we point out that, for a spherically
symmetric interaction potential, the interparticle stress tensor is
symmetric, $ \sigma_{xy}^{\rm int} = \sigma_{yx}^{\rm int} $. Furthermore, for a
spherically symmetric external potential, we also have
$ \sum_{i=1}^NF_{ix}^{\rm ext}y_i=\sum_{i=1}^NF_{iy}^{\rm
  ext}x_i$ and the terms $\cos\varphi_i(t')y_i(t')$ and
$\sin\varphi_i(t')x_i(t')$ in Eq.~\eqref{eq:LRAF2} yield identical
contributions, so that symmetrization of $ \widetilde{\sigma} $ is not
necessary. We thus have in this case
\begin{align}
\notag {\langle A(t) \rangle}_{\rm st}^{(\dot{\gamma})} & -  {\langle A \rangle}_{\rm st}= \frac{\dot\gamma \mu_{\rm t}}{2 D_{\rm t}} \int_0^t dt'\left\langle A(t)\widetilde{\sigma}_{xy}(t')\right\rangle_{\rm st}\\
+\frac{\dot{\gamma}}{4D_{\rm t}}&{\left\langle A(t)\sum_{i=1}^N\left[{x}_i(t)y_i(t)  - {x}_i(0)y_i(0) \right]\right\rangle}_{\rm st}.
\label{eq:LRAF_final}
\end{align}
Formula~\eqref{eq:LRAF_final} is the most important result of this
paper. While Eq.~\eqref{eq:LR_final} does not require the mentioned symmetries,
Eq.~\eqref{eq:LRAF_final} is significantly simpler, because it
contains only state variables. Moreover, it contains no response to shear rotation. Note that, in addition to the specific
symmetry of the considered system and the observable $ A $,
Eq.~\eqref{eq:LRAF_final} also requires that the system is in steady state before the shear flow is applied, as indicated.

The response formula~\eqref{eq:LRAF_final} may be interpreted as a
generalized Green-Kubo relation for interacting ABPs subject to an
external potential.  It differs from the traditional (equilibrium)
Green-Kubo relation~\cite{Kubo1991, Fuchs2005} in two ways: the stress
tensor $ \sigma_{xy}^{\rm int} $ is replaced by the generalized one
$\widetilde{\sigma}_{xy}$, and there is the second term in
Eq.~\eqref{eq:LRAF_final} that is present because of the breaking of
detailed balance in the unperturbed system.

The use of Eq.~\eqref{eq:LRAF_final}
for unconfined systems is unclear at the moment, because $x_iy_i$
grows unboundedly with the system size. While we use
Eq.~\eqref{eq:LRAF_final} to compute the response in confined
geometries in this paper (see the examples below), investigating its applicability in bulk is an important topic for future research.

Finally, we note that the details of the stochastic process underlying the angle $\varphi$ given in Eq.~\eqref{eq:LE3} do not appear explicitly in Eq.~\eqref{eq:LRAF_final}. We will explore this observation in the next subsection, thereby finding that the presented scheme is readily applied to a more general setup, yielding Eq.~\eqref{eq:LRAF_finalGeneral} below.

\subsection{Three space dimensions and more general setups}
\label{subsec:General}
While we have so far considered two spatial dimensions, we aim here to derive the analog of Eq.~\eqref{eq:LRAF_final} for more general setups of spherical particles.  
This is inspired by the mentioned observation that $ D_{\rm r} $ and $ \mu_{\rm r} $ are absent in formula~\eqref{eq:LRAF_final}, suggesting its independence of the details of the dynamics of the swim velocity vector.

We start with the multi-dimensional Langevin equation for position $ \vct{r}_i =(x_i,y_i,\dots)^T$ of particle $ i $, 
\begin{equation}
	\dot{\vct{r}}_i = {\boldsymbol \kappa}\cdot \vct{r}_i + \vct{u}_i + \mu_{\rm t}\vct{F}^{\rm int}_i + \mu_{\rm t}\vct{F}^{\rm ext}_i + \mu_{\rm t}\vct{f}_i,
\label{eq:LEsGeneral}
\end{equation}
where $ \boldsymbol \kappa = \dot{\gamma} \hat{\bf x}\otimes\hat{\bf y}$, given in terms of unit vectors, is the shear-velocity tensor. $ \vct{u}_i $ is the swim velocity, and Gaussian white noises $ \vct{f}_i $ satisfy
\begin{equation}
\langle \vct{f}_i(t) \rangle = 0, \ \ \ \langle \vct{f}_i(t) \otimes \vct{f}_j(t') \rangle = \frac{2D_{\rm t}}{\mu_{\rm t}^2}\mathbb{I}\delta_{ij}\delta(t-t'),
\label{noise_properties3D}
\end{equation}
with $ \otimes $ denoting the tensor product and $ \mathbb{I} $ being identity matrix.

The swim velocity $ \vct{u}_i $ obeys its own stochastic process. For the following arguments to be valid, we require the process for $ \vct{u}_i $ to be random and unbiased in the absence of shear, and its stochastic properties (e.g. its noise) uncorrelated with the noise $\vct{f}_i$ in Eq.~\eqref{noise_properties3D}. Furthermore, with shear, $\vct{u}_i$ may be subject to a shear-torque (compare Eq.~\eqref{eq:LE3}). When superposing two shear flows, $ \boldsymbol \kappa = \dot{\gamma} (\hat{\bf x}\otimes\hat{\bf y}+\hat{\bf y}\otimes\hat{\bf x})$, as done in Subsec.~\ref{subsec:LR_another_form}, this shear-torque drops out.
As a specific example, adding shear to ABPs in three spatial dimensions as given in Ref.~\cite{Sharma2016}, these conditions are naturally met, and Eq.~\eqref{eq:LRAF_finalGeneral} below is valid. 

	The total action $ \mathcal{A} $ of the system can be written as the sum of the action $ \mathcal{A}_{\rm r} $ following from Eq.~\eqref{eq:LEsGeneral}, and $\mathcal{A}_{\rm u} $ deduced from the swim velocity $\vct{u}_i $,
\begin{equation}
	\mathcal{A}(t, \dot{\gamma}) = \mathcal{A}_{\rm r}(t, \dot{\gamma}) + \mathcal{A}_{\rm u}(t, \dot{\gamma}).
\label{eq:action_sum}
\end{equation}
These parts are additive if the noise for $ \vct{r}_i $ in Eq.~\eqref{noise_properties3D} is uncorrelated with the process of $ \vct{u}_i $.
When superposing the mentioned two shear directions, the dependence of $ \mathcal{A}_{\rm u} $ on shear rate drops out. 
Performing the perturbation of $\mathcal{A}_{\rm r}(t, \dot{\gamma})$ and following the procedures described in Subsec.~\ref{subsec:LR_another_form}, one obtains the form of Eq.~\eqref{eq:LRAF_final}  with $ v_0\cos{\varphi_i(t')} $ replaced by 
$ u_{ix}(t') $. Specifically,
\begin{align}
\notag {\langle A(t) \rangle}_{\rm st}^{(\dot{\gamma})} & -  {\langle A \rangle}_{\rm st}= \frac{\dot\gamma \mu_{\rm t}}{2 D_{\rm t}} \int_0^t dt'\left\langle A(t)\widetilde{\sigma}_{xy}(t')\right\rangle_{\rm st}\\
+\frac{\dot{\gamma}}{4D_{\rm t}}&{\left\langle A(t)\sum_{i=1}^N\left[{x}_i(t)y_i(t)  - {x}_i(0)y_i(0) \right]\right\rangle}_{\rm st},
\label{eq:LRAF_finalGeneral}
\end{align}
where the swim stress tensor in $ \widetilde{\sigma}_{xy} $ is given by $\sigma^{\rm s}_{xy} = -\sum_{i = 1}^N \frac{1}{\mu_t}u_{ix}  y_i $.

\section{Analytical Examples}
\label{sec:Analytical_Examples}
The main purpose of this section is to demonstrate the use of the response formulas~\eqref{eq:LR_final} and~\eqref{eq:LRAF_final} for solvable analytical cases, namely for a single active particle
in free space or confined by a harmonic potential. Furthermore, we
complement the previously known results, providing the transient
regime of the response computed in
Subsec.~\ref{subsec:Active_particle_hp} as well as correlation
functions in Appendix~\ref{subapp:AppendixRight}. Throught this section, we consider a two-dimensional system and set the torque $ M = 0 $ and the rotational mobility $ \mu_{\rm r} = 1 $ for 
simplicity.

\subsection{Free active particle}
\label{subsec:Free_Active_Particle}
We apply here Eq.~\eqref{eq:LR_final} to compute the response to shear
of the mean displacement
$ {\langle x(t) \rangle}_0^{(\dot{\gamma})} - {\langle x(t) \rangle}_0
$ for a single free self-propelled particle, and show that the
response formula reproduces the result of Ref.~\cite{tenHagen2011},
directly computed in the sheared system.

Setting $ A = x $ and $ N = 1 $ in Eq.~\eqref{eq:LR_final}, and using
Eq.~\eqref{eq:LEs} to rewrite Eq.~\eqref{eq:LR_final} in terms of
random force and torque, one has
\begin{align}
& \notag {\langle x(t) \rangle}_0^{(\dot{\gamma})} - {\langle x(t) \rangle}_0\\
& = \frac{\dot{\gamma}}{4}\int_0^tdt'\Big\langle \frac{2\mu_{\rm t}}{D_{\rm t}}x(t)f_{x}(t')y(t') - \frac{1}{D_{\rm r}}x(t)g(t')\Big\rangle_0.
\label{eq:mean_x_initial}
\end{align}
Given the initial condition
$ \Gamma(0) = \{x(0), y(0), \varphi(0)\} $, the correlation functions
appearing in Eq.~(\ref{eq:mean_x_initial}) can be calculated
explicitly:
\begin{subequations}\label{eq:mean_x_correlators}
\begin{align}
& \notag {\langle x(t)f_x(t')y(t') \rangle}_0\\
& = \frac{2D_{\rm t}}{\mu_{\rm t}}\left[y(0) + \frac{v_0\sin\varphi(0)}{D_{\rm r}}\left(1-e^{-D_{\rm r}t'}\right)\right],
\label{eq:mean_x_first_correlator}
\end{align}
\begin{equation}
{\langle x(t)g(t') \rangle}_0 = 2v_0\sin\varphi(0)\left[e^{-D_{\rm r}t} - e^{-D_{\rm r}t'}\right].
\label{eq:mean_x_second_correlator}
\end{equation}
\end{subequations}
Finally, performing the time integral in Eq.~\eqref{eq:mean_x_initial}
gives
\begin{align}
\notag {\langle x(t) \rangle}_0^{(\dot{\gamma})} & - {\langle x(t) \rangle}_0 = \dot{\gamma}\Bigg\{y(0)t +\frac{v_0\sin\varphi(0)}{D_{\rm r}}\\
\times \Bigg[&t\left(1-\frac{1}{2}e^{-D_{\rm r}t}\right) -\frac{1}{2D_{\rm r}}\left(1-e^{-D_{\rm r}t}\right)\Bigg]\Bigg\}.
\label{eq:mean_x_final}
\end{align}
This reproduces Eq.~(11) in Ref.~\cite{tenHagen2011}, which, as
mentioned, was computed without use of the response theory.

\subsection{Active particle in a harmonic trap}
\label{subsec:Active_particle_hp}
We compute similarly the response to shear of a single active particle in
a harmonic potential $ U^{\rm ext} = \frac{k}{2}(x^2+y^2) $. The
system is assumed to be in steady state before shear is applied. For
concreteness, we focus on computing the observable $ A(t)= x(t)y(t) $,
which characterizes the shape of the density distribution. We compute independently the left and
right-hand sides of Eq.~\eqref{eq:LRAF_final} thereby verifying the equation explicitly. The stationary limit of the response was studied
before in Ref.~\cite{Li2017} and agrees with our findings.

For this system, the Langevin equations~\eqref{eq:LEs} reduce to
\begin{subequations}\label{eq:AE_LEs}
\begin{align}
&\dot{x} = \dot{\gamma}y + v_0\cos\varphi - \mu_{\rm t} kx + \mu_{\rm t} f_{x} \label{eq:AE_LE1},\\
&\dot{y} = v_0\sin\varphi - \mu_{\rm t} ky + \mu_{\rm t} f_{y} \label{eq:AE_LE2},\\
&\dot{\varphi}= -\frac{\dot{\gamma}}{2} + g \label{eq:AE_LE3}.
\end{align}
\end{subequations}
The corresponding solutions for $ t \geq 0 $ read
\begin{subequations}\label{eq:AE_Sol}
\begin{align}
\notag x(t) = \ & \dot{\gamma}e^{-\mu_{\rm t} kt}\int_0^tdse^{\mu_{\rm t} ks}y(s)\\
\notag & + v_0e^{-\mu_{\rm t} kt}\int_{-\infty}^tdse^{\mu_{\rm t} ks}\cos\varphi(s)\\
& + \mu_{\rm t} e^{-\mu_{\rm t} kt}\int_{-\infty}^tdse^{\mu_{\rm t} ks}f_x(s),
\label{eq:AE_Sol1}
\end{align}
\begin{align}
\notag y(t) = \ & v_0e^{-\mu_{\rm t} kt}\int_{-\infty}^tdse^{\mu_{\rm t} ks}\sin\varphi(s)\\
& + \mu_{\rm t} e^{-\mu_{\rm t} kt}\int_{-\infty}^tdse^{\mu_{\rm t} ks}f_y(s),
\label{eq:AE_Sol2}
\end{align}
and
\begin{equation}
\varphi(t) = \varphi(-\infty) -\frac{\dot{\gamma}}{2}t + \int_{-\infty}^tdsg(s).
\label{eq:AE_Sol3}
\end{equation}
\end{subequations}
For $ t < 0 $, the first term on the right-hand side of
Eq.~\eqref{eq:AE_Sol1} and the second term on the right-hand side of
Eq.~\eqref{eq:AE_Sol3} are absent.  Since $ \varphi(t) $ never reaches 
stationary state, this variable depends on the initial condition
$ \varphi(-\infty) $. However, any observable $ A(x(t), y(t)) $
depending only on the coordinates remains \textit{independent} of
$ \varphi(-\infty) $ in both the unperturbed stationary state and the
perturbed transient and stationary regimes, since this initial
condition is forgotten for stationary values of $ x(t) $ and
$ y (t) $.  For the observable $ A(t)= x(t)y(t) $,
Eq.~\eqref{eq:LRAF_final} reads explicitly
\begin{align}
\notag &{\langle x(t)y(t) \rangle}_{\rm st}^{(\dot{\gamma})} - {\langle xy \rangle}_{\rm st}\\
\notag & = \frac{\dot{\gamma}}{4D_{\rm t}}{\left\langle x(t)y(t)\left[x(t)y(t)-x(0)y(0)\right]\right\rangle}_{\rm st}\\
\notag & - \frac{\dot{\gamma}v_0}{2D_{\rm t}}\int_0^tdt'{\left\langle x(t)y(t)\cos\varphi(t')y(t')\right\rangle}_{\rm st}\\
&+\frac{\dot{\gamma}\mu_{\rm t} k}{2D_{\rm t}}\int_0^tdt'{\left\langle x(t)y(t)x(t')y(t')\right\rangle}_{\rm st} .
\label{eq:AE_equality}
\end{align}
In the following, we verify Eq.~(\ref{eq:AE_equality}) by
independently computing both sides.

We start by computing the left-hand side of Eq.~\eqref{eq:AE_equality}
up to linear order in shear rate $ \dot{\gamma} $. We provide here
only the final result (computational details can be found in
Appendix~\ref{subapp:AppendixLeft}),
\begin{align}
\notag {\langle x(t)y(t) \rangle}_{\rm st}^{(\dot{\gamma})} &  = \frac{\dot{\gamma}D_{\rm t}}{2{(\mu_{\rm t} k)}^2}\left(1 - e^{-2 \mu_{\rm t} k t}\right) \\
\notag & + \frac{\dot{\gamma}v_0^2}{4{(\mu_{\rm t} k)}^2\left[D_{\rm r}^2-{(\mu_{\rm t} k)}^2\right]}\Bigg\{D_{\rm r}\left(1-e^{-2\mu_{\rm t} kt}\right) \\
& - \frac{2{(\mu_{\rm t} k)}^2}{D_{\rm r}+\mu_{\rm t} k}\left(1-e^{-(D_{\rm r}+\mu_{\rm t} k)t}\right)\Bigg\}.
\label{eq:AE_equality2}
\end{align}
Here, the first term is a passive contribution, while the second term is due to activity (note the
presence of $ v_0 $ and $ D_{\rm r} $ and the absence of
$ D_{\rm t} $). The term $ {\langle xy \rangle}_{\rm st} $ in
Eq.~\eqref{eq:AE_equality} vanishes by symmetry.

We then compute independently the right-hand side of
Eq.~\eqref{eq:AE_equality} in Appendix~\ref{subapp:AppendixRight}, and
find it to be identical to Eq.~\eqref{eq:AE_equality2}. This verifies
explicitly the validity of the response
relation~\eqref{eq:AE_equality}.

We close this subsection with a discussion of the physics contained in
Eq.~\eqref{eq:AE_equality2}. First, one can show that both terms
($ \propto D_{\rm t} $ and $ \propto v_0^2 $) in
Eq.~\eqref{eq:AE_equality2} are nonnegative, indicating that the
activity increases the positive response of a passive particle and
that the total response is not negative. This means that, although
$ {\langle x(t) \rangle}_{\rm st}^{(\dot{\gamma})}= {\langle y(t)
  \rangle}_{\rm st}^{(\dot{\gamma})} = 0 $, we have
$ {\langle x(t)y(t) \rangle}_{\rm st}^{(\dot{\gamma})} \geq 0 $,
because the shear flow couples the two directions and breaks the
isotropicity of the system. As a result of the shear flow, the
particle tends to be in the coordinate quadrants where $ x $ and $ y $
are either both positive or both negative. Also note that passive and
active contributions enter the response~\eqref{eq:AE_equality2}
independently, i.e., translational diffusion and activity are not
coupled (there are no terms containing both $ D_{\rm t} $ and
$ v_0^2 $). We think that this could be a feature of \textit{linear}
response, while in the nonlinear case one may observe a coupling
between these two contributions.

\begin{figure}[!t]
\includegraphics[width=1.0\linewidth]{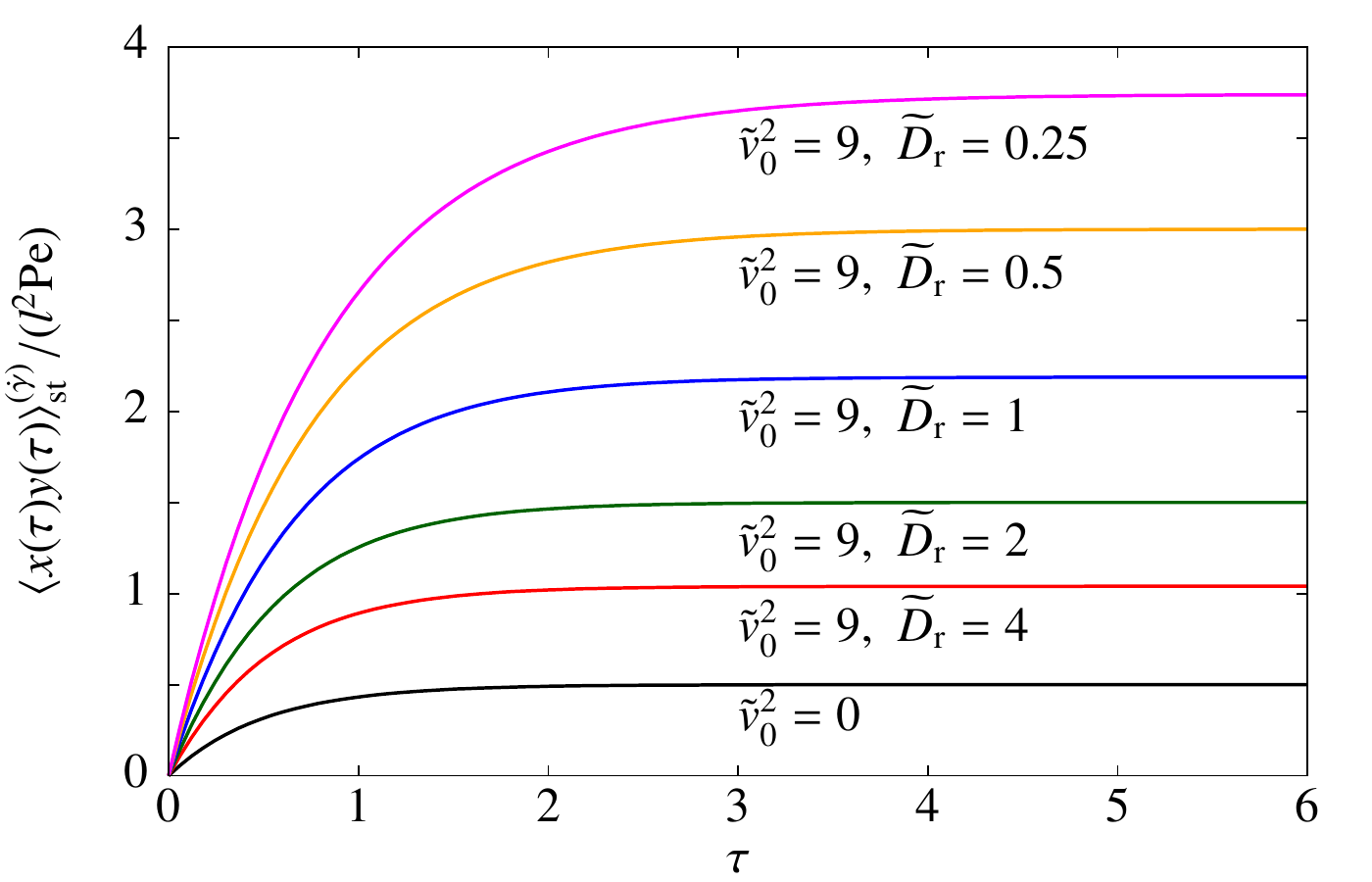}
\caption{\label{fig:aparticle_in_hp_response} Rescaled response,
  given by Eq.~\eqref{eq:AE_equality2_plot}, of a single active
  particle in a harmonic potential to shear flow as a function of
  rescaled time $ \tau $ after the flow is applied. The results are
  given for different values of $ {\tilde{v}_0}^2 $ and $ \widetilde{D}_{\rm r} $ (the case $ {\tilde{v}_0}^2  = 0 $
  corresponds to a passive particle).}
\end{figure}

In order to visualize Eq.~\eqref{eq:AE_equality2}, we rewrite it in
terms of dimensionless parameters: $ \tau \equiv \mu_{\rm t} k t $ (describing time in units of the relaxation time $ \frac{1}{\mu_{\rm t} k} $ of the trap), $ {\rm Pe} \equiv \frac{\dot{\gamma}}{\mu_{\rm t} k} $ 
(P\'eclet number), $ \widetilde{D}_{\rm r} \equiv \frac{D_{\rm r}}{\mu_{\rm t} k} $ (normalizing rotational relaxation time $D_{\rm r}^{-1}$ by the relaxation time $ \frac{1}{\mu_{\rm t} k} $ of the trap), $ {\tilde{v}_0}^2 \equiv \frac{v_0^2}{D_{\rm t}\mu_{\rm t} k} $ (comparing swim speed to translational diffusion and the strength of the trap). 
Rescaling $ {\langle x(t)y(t) \rangle}_{\rm st}^{(\dot{\gamma})} $ by the unit
of squared length, $ l^2 \equiv \frac{D_{\rm t}}{\mu_{\rm t} k} $, and
dividing by $ {\rm Pe} $, we rewrite Eq.~\eqref{eq:AE_equality2} as
\begin{align}
\notag &\frac{{\langle x(\tau)y(\tau) \rangle}_{\rm st}^{(\dot{\gamma})}}{l^2 {\rm Pe}} = \frac{1}{2}\left(1 - e^{-2 \tau}\right) + \frac{{\tilde{v}_0}^2}{4\left({\widetilde{D}_{\rm r}}^2-1\right)}\\
&\times\Bigg\{\widetilde{D}_{\rm r}\left(1-e^{-2\tau}\right) - \frac{2}{\widetilde{D}_{\rm r}+1}\left(1-e^{-\left(\widetilde{D}_{\rm r}+1\right)\tau}\right)\Bigg\}.
\label{eq:AE_equality2_plot}
\end{align}
This function is plotted in Figure~\ref{fig:aparticle_in_hp_response}
as a function of $ \tau $ for different values of $ {\tilde{v}_0}^2 $ and $ \widetilde{D}_{\rm r} $,
thereby summarizing the above discussions. One can see in
Fig.~\ref{fig:aparticle_in_hp_response} that the response increases as $ \widetilde{D}_{\rm r} $ decreases, because the active motion becomes more persistent.

\section{Numerical Example: Interacting particles in two space dimensions}
\label{sec:Numerical_Example}
The potential utility of Eq.~\eqref{eq:LRAF_final} lies in its
application to experiments and computer simulations of interacting
particles, which we address in this section. We demonstrate this
numerically for a two-dimensional system of particles trapped in a harmonic potential
$U^{\rm ext} = \frac{k}{2}\sum_{i=1}^N (x_i^2+y_i^2)$, and interacting
with a short-ranged harmonic repulsion,
$U_{ij}^{\rm int}(r_{ij})=\frac{k^{\rm int}}{2} (r_c-r_{ij})^2$ for
$r_{ij}<r_c$ (where $ r_{ij} $ is the distance between particle $i$
and particle $j \neq i$) and $U_{ij}^{\rm int}=0$ otherwise. 
A similar scenario has been studied in Ref.~\cite{Warren2012} for passive particles.

We take the radius of interaction $r_c=1$ as our space unit, and
choose $k=1$ and the mobility $\mu_{\rm t}=1$, thus fixing the time and energy
scales. The dynamics, Eq.~\eqref{eq:LEs}, is integrated using Euler
time-stepping.

\begin{figure}[!t]
\includegraphics[width=0.8\linewidth]{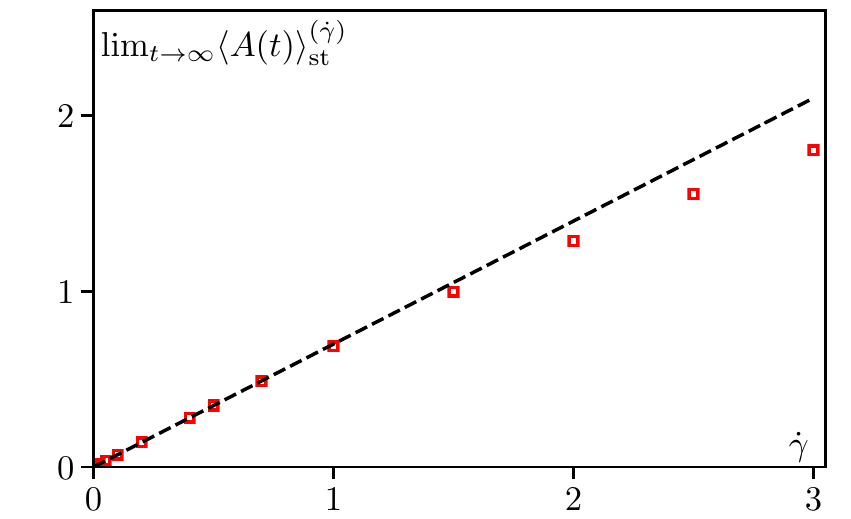}\\
\includegraphics[width=0.8\linewidth]{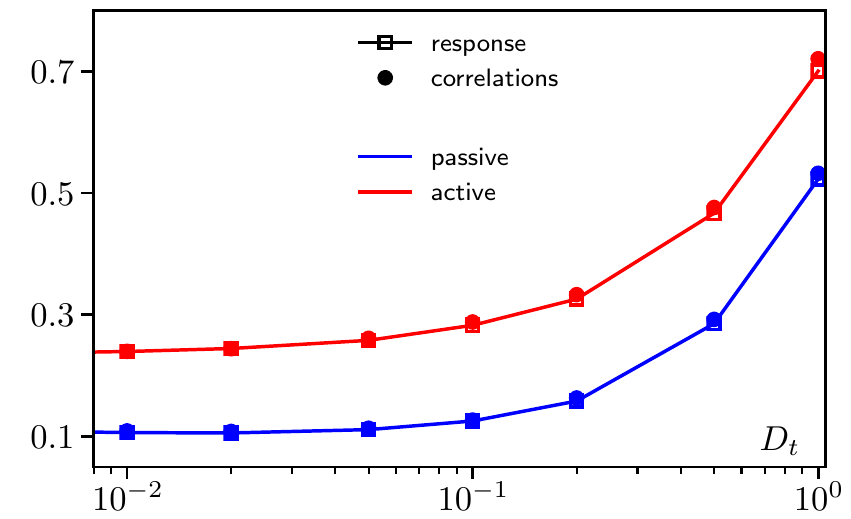}\\
\includegraphics[width=0.8\linewidth]{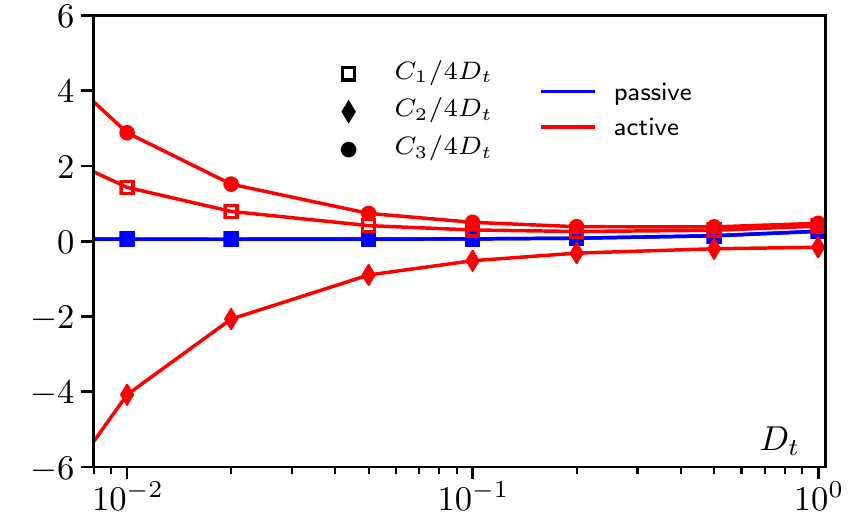}
\caption{
\label{fig:harmonic-pot}
Numerical results for a suspension of interacting active
particles ($ M_i = 0 $). {\bf Top:} Response measured in the sheared system for
active particles with $D_{\rm t}=1$. The dashed black line is a fit in
the linear regime at small $\dot\gamma$. {\bf Center:} Comparison of
the response and correlations, i.e., the left-hand and right-hand
sides of Eq.~\eqref{eq:LRAF_interms}, respectively. {\bf Bottom:}
Detail of each term on the right-hand side of
Eq.~\eqref{eq:LRAF_interms}. Parameters: $N=10$ particles, $k=1$,
$k^{\rm int}=0.5$, $\mu_{\rm t}=1$, and $v_0=D_{\rm r}=\mu_{\rm r}=1$ for active
particles. Time step $dt=0.02$.}
\end{figure}

We measure independently the two sides of Eq.~\eqref{eq:LRAF_final}
for $ A(t) = \sum_{i=1}^Nx_i(t)y_i(t) $ in steady state
($ t \to \infty $) which characterizes the distortion of the density
distribution due to the shear flow, and which is identified with $\sigma_{xy}^{\rm ext}$ in Eq.~\eqref{eq:pseudo_stress}, $ A(t) = \sum_{i=1}^Nx_i(t)y_i(t) =\sigma_{xy}^{\rm ext}$. The boundary term
$ {\left\langle A(t)\sum_{i=1}^Nx_i(0)y_i(0) \right\rangle}_{\rm st}$
in Eq.~\eqref{eq:LRAF_final} is then irrelevant. It is illustrative to
split Eq.~(\ref{eq:LRAF_final}) into its different contributions,
\begin{equation}
\frac{{\langle A(t) \rangle}_{\rm st}^{(\dot{\gamma})} - {\langle A \rangle}_{\rm st}}{\dot\gamma}  = \frac{C_1+C_2+C_3}{4 D_{\rm t}},
\label{eq:LRAF_interms}
\end{equation}
where
\begin{subequations}\label{eq:C}
\begin{align}
&C_1 ={\left\langle A(t)\sum_{i=1}^N x_i(t)y_i(t)\right\rangle}_{\rm st},\label{eq:C1}\\
&C_2 =-2v_0\int_0^tdt' {\left\langle A(t)\sum_{i=1}^N\cos\varphi_i(t')y_i(t')\right\rangle}_{\rm st},\label{eq:C2}\\
&C_3 =2 \mu_{\rm t} \int_0^tdt' {\left\langle A(t)\left[\sigma^{\rm int}_{xy}(t')-\sum_{i=1}^NF_{ix}^{\rm ext}y_i(t')\right]\right\rangle}_{\rm st}.\label{eq:C3}
\end{align}
\end{subequations}
As in the previous section, due to the isotropicity of the unsheared
system, $ {\langle A \rangle}_{\rm st} = 0 $. 

The results obtained for $N=10$ particles interacting with a spring constant $k^{\rm int}=0.5$
for various $D_{\rm t}$ are shown in Fig.~\ref{fig:harmonic-pot} for
both active (with $v_0=D_{\rm r}=\mu_{\rm r}=1$) and passive ($v_0=0$)
particles. The response is first obtained by simulating the sheared
system at different shear rates and extracting the small $\dot\gamma$
behavior as shown in Fig.~\ref{fig:harmonic-pot}~(top). In
Fig.~\ref{fig:harmonic-pot}~(center) we then compare to the right-hand
side of Eq.~(\ref{eq:LRAF_interms}), obtained by measuring the
appropriate correlation functions in the unperturbed system. We find
the two measurements to agree perfectly, given the numerical
uncertainty.

First, Fig.~\ref{fig:harmonic-pot}~(center) shows that the response is
positive and is increased by activity, as was observed for a single
particle in the previous section. Second, it is interesting to compare
the limits $D_{\rm t}\to 0$ in the passive and active cases because
they show qualitatively different behaviors. In the passive case, this
corresponds to the zero-temperature limit so that the system becomes
frozen in a minimal energy configuration. We find numerically that
both $C_1$ and $C_3$ in Eqs.~\eqref{eq:C1} and~\eqref{eq:C3} are
proportional to $D_{\rm t}$ in this limit ($C_2$ vanishes for passive
particles). As a result, the two terms $\frac{C_1}{4D_{\rm t}}$ and
$\frac{C_3}{4D_{\rm t}}$ become constant at small $D_{\rm t}$, as
shown in Fig.~\ref{fig:harmonic-pot} (bottom). In contrast, active
particles are still moving even at $D_{\rm t}=0$ so that the
correlators $C_i$ do not vanish. As a result, each of the terms on the
right-hand side of Eq.~\eqref{eq:LRAF_interms} diverges when
$D_{\rm t}\to 0$ in such a way that the sum remains constant. While
confirming the applicability of Eq.~(\ref{eq:LRAF_final}) to many-body
systems, this analysis also highlights a limitation of our
formulation. Indeed, our derivation necessitates a finite $D_{\rm t}$,
since the distributional description of particle trajectories relies
on the presence of stochasticity. However, $D_{\rm t}$ is often
negligible in active systems, since the particles' motion in that case
is primarily due to activity~\cite{Deseigne2010, Ginot2015,
  Bicard2013}.  It may thus be desirable to obtain formulas valid for
$D_{\rm t}=0$, as is done in Ref.~\cite{Szamel2017} for a different
active particle model.


\begin{figure}[!t]
\includegraphics[width=1.0\linewidth]{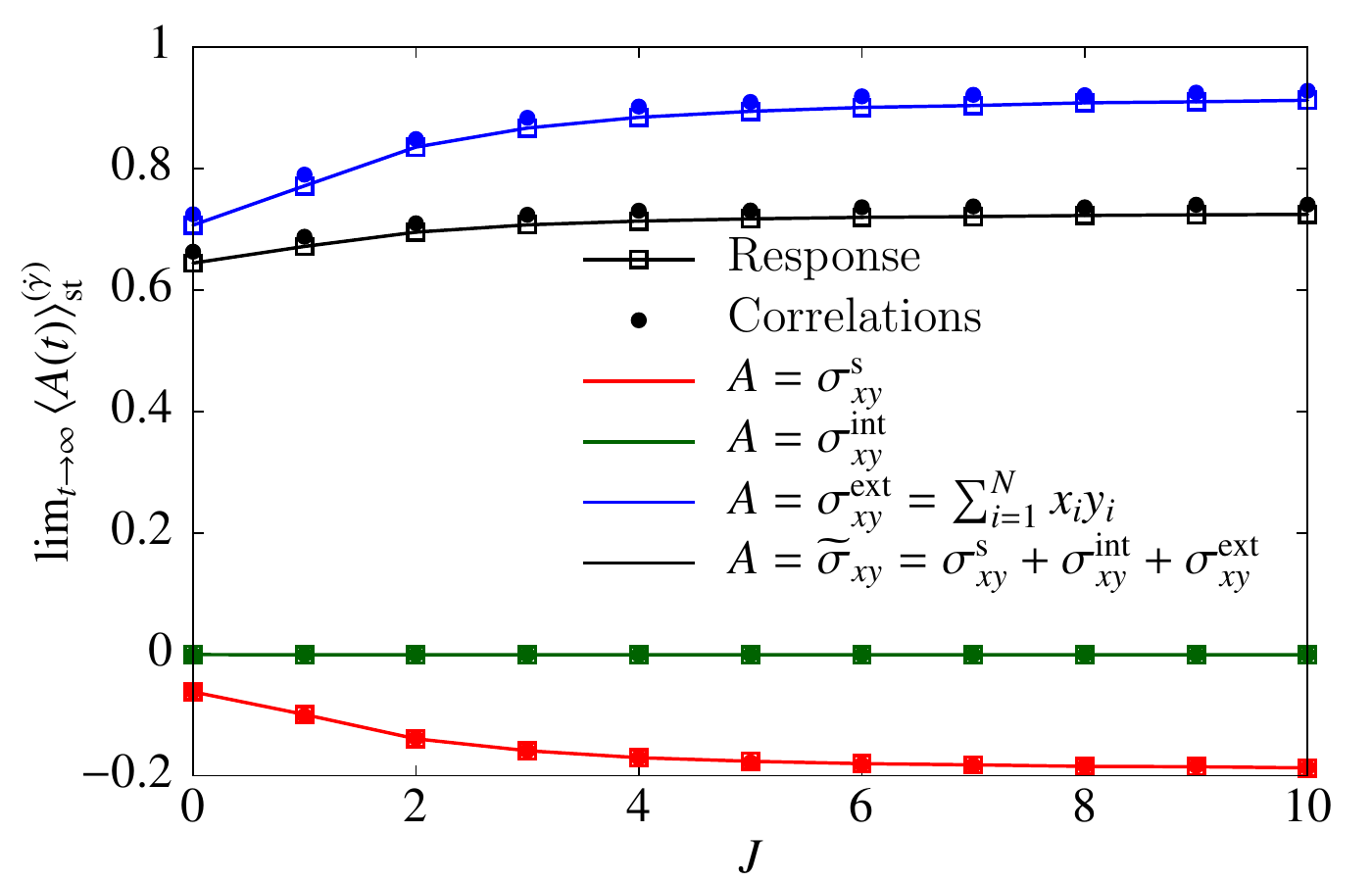}
\caption{\label{fig:alignment}Numerical results for a suspension of interacting active particles subject to alignment interactions given in Eq.~\eqref{eq:alignment}, as a function of strength of alignment $J$. Shown is the response of the terms appearing in the pseudostress tensor $ \widetilde{\sigma}_{xy} $ defined in Eq.~\eqref{eq:pseudo_stress}: the swim stress tensor $ \sigma^{\rm s}_{xy} $, the conventional interaction stress tensor $ \sigma_{xy}^{\rm int} $, and the external stress tensor $ \sigma^{\rm ext}_{xy}$. The latter equals the observable studied in Fig.~\ref{fig:harmonic-pot}, i.e., 
$ \sigma^{\rm ext}_{xy} = \sum_{i=1}^Nx_iy_i $. Parameters as in Fig.~\ref{fig:harmonic-pot} and 
$ D_{\rm t} = 1 $.}
\end{figure}

Next, we include alignment interactions modeled by torques 
\begin{equation}
M_i = -J \sum_{j = 1 (j \neq i)}^N\sin(\varphi_i-\varphi_j),
\label{eq:alignment}
\end{equation}
where the sum runs over particles in contact, i.e., with the interparticle distance $ r_{ij} < r_c $.
This torque arises from a typical ``spin''-interaction, formed by scalar products of particle orientation vectors $ \hat{\vct{u}}_i $, and is hence symmetric under interchange of $ x $ and $ y $ coordinates. Formula~\eqref{eq:LRAF_final} can therefore be applied.

Additionally to $ \sigma_{xy}^{\rm ext} $, we compute also $ \sigma_{xy}^{\rm int} $ and $ \sigma_{xy}^{\rm s} $  defined in 
Eq.~\eqref{eq:pseudo_stress}. The results are given in Fig.~\ref{fig:alignment}. First, we note that the magnitudes of all stress tensor components increase with the alignment.  We may expect that strong alignment renders the particle cloud into an elongated shape, which is more susceptible to shear.  The saturation for large values of $J$ is expected, as, once all velocities are perfectly aligned, increasing $J$ has no effect.
Notably, the interaction stress $\sigma_{xy}^{\rm int}$ remains zero within errors. While this appears plausible, in the sense that the external force balances the shear force in Eq.~\eqref{eq:LEs}, we are not aware of a proof that $\sigma_{xy}^{\rm int}=0$ exactly. 





\section{Conclusion}
\label{sec:Conclusion}
In this paper, we have studied the linear response to simple shear flow of interacting active Brownian
particles with external forces. The path integral formalism yields, in two space dimensions, the linear response formula~\eqref{eq:LR_final} relating any
time dependent state observable of the sheared system to correlation
functions of the unsheared system.

For systems and observables obeying $ xy $ symmetry, the
initial response formula was shown to simplify such that the final
result, Eq.~\eqref{eq:LRAF_finalGeneral}, contains only state variables and is valid in any space dimension and for a wider set of activity models. This simplification is a
consequence of the fact that shear in the $ x $ direction and shear in
the $ y $ direction, having opposite torques, are equivalent for
$ xy $ symmetric systems and observables. This form of the response
formula is particularly advantageous since it involves quantities that
are typically easier to measure. 

Next, we investigated the morphology and stresses of a two-dimensional cluster formed by $ N $
interacting active particles confined by a harmonic potential under
shear. Performing analytical computations for $ N = 1 $ and numerical
simulations for $ N > 1 $ particles, we found that the average of
$ \sum_{i=1}^Nx_iy_i $ under shear is nonnegative and larger compared
to passive particles. We also found that increasing the persistence of active particles (decreasing $D_{\rm r}$) or adding alignment interactions between the particles
increases the response to shear, so that the magnitudes of the found stresses increase.

Future work may consider the limit of zero translational diffusion, as
well as the viscosity of a suspension of active Brownian particles.
Finally, the extension to higher order responses is also a promising
avenue to explore, since this could, for example, shed light on the
coupling between shear translation and shear rotation.

\begin{acknowledgments}
We thank U.~Basu, B.~ten~Hagen, U.~S.~Schwarz, G.~Szamel, and Th.~Voigtmann for valuable discussions. K.~Asheichyk and M.~Kr\"uger were supported by Deutsche Forschungsgemeinschaft (DFG) Grant 
No. KR 3844/2-2. K.~Asheichyk also acknowledges Studienstiftung des deutschen Volkes, the Physics Department of the University of Stuttgart, and S.~Dietrich for their support. 
C.~M.~Rohwer acknowledges support by S.~Dietrich. 
\end{acknowledgments}


\onecolumngrid

\appendix
\section{Detailed computation of the response in Subsec.~\ref{subsec:Active_particle_hp}}
\label{app:Appendix}

This Appendix sets out the necessary steps to compute all terms of Eq.~\eqref{eq:AE_equality} explicitly.

\subsection{Computation of the left-hand side of Eq.~\eqref{eq:AE_equality}}
\label{subapp:AppendixLeft}
First, one can show that
\begin{equation}
{\langle \cos\varphi(s_1) \cos\varphi(s_2) \rangle}_{\rm st}^{(\dot{\gamma})} = {\langle \sin\varphi(s_1)\sin\varphi(s_2) \rangle}_{\rm st}^{(\dot{\gamma})} = \frac{1}{2}\cos\left\{\frac{\dot{\gamma}}{2}\left[s_1-s_2\right]\right\}e^{-D_{\rm r}|s_1-s_2|}
\label{eq:AE_coscos_sinsin}
\end{equation}
and
\begin{equation}
{\langle \cos\varphi(s_1) \sin\varphi(s_2) \rangle}_{\rm st}^{(\dot{\gamma})} = \frac{1}{2}\sin\left\{\frac{\dot{\gamma}}{2}\left[s_1-s_2\right]\right\}e^{-D_{\rm r}|s_1-s_2|},
\label{eq:AE_cossin}
\end{equation}
in agreement with Ref.~\cite{Li2017}. For the unsheared correlators, we hence have
\begin{equation}
{\langle \cos\varphi(s_1)\cos\varphi(s_2) \rangle}_{\rm st} = {\langle \sin\varphi(s_1)\sin\varphi(s_2) \rangle}_{\rm st} =\frac{1}{2}e^{-D_{\rm r}|s_1-s_2|}
\label{eq:AE_coscos_sinsin_unsheared}
\end{equation}
and
\begin{equation}
{\langle \cos\varphi(s_1)\sin\varphi(s_2) \rangle}_{\rm st} = 0.
\label{eq:AE_cossin_unsheared}
\end{equation}
Results~\eqref{eq:AE_coscos_sinsin} --~\eqref{eq:AE_cossin_unsheared} do not depend on the initial angle $ \varphi(-\infty) $, because we consider stationary correlation functions, i.e., we let the angle to evolve from far away in the past  [$ -\infty $ limit in Eq.~\eqref{eq:AE_Sol3}] such that the initial angle is forgotten. We note that this limit does not commute with the limit $ D_{\rm r} \to 0 $. This is physical, because for times smaller than $ \frac{1}{D_{\rm r}} $ a particle remembers its initial orientation.

Due to Eq.~\eqref{eq:AE_cossin_unsheared},
\begin{equation}
{\langle xy \rangle}_{\rm st} = 0,
\label{eq:AE_equality1}
\end{equation}
as in the case of a trapped passive particle. This result is intuitive, because the unsheared system is $ xy $ symmetric and the particle moves around the origin. For $ {\langle x(t)y(t) \rangle}_{\rm st}^{(\dot{\gamma})} $ linear 
in $ \dot{\gamma} $, the relevant nonzero correlators are those given by Eqs.~\eqref{eq:AE_coscos_sinsin} and~\eqref{eq:AE_cossin} linear in $ \dot{\gamma} $ and 
$ \langle f_y(s_1)f_y(s_2) \rangle = \frac{2D_{\rm t}}{\mu_{\rm t}^2}\delta(s_1-s_2) $. The contribution of correlator~\eqref{eq:AE_cossin} is, however, zero due to the symmetry of time integrals containing it. Multiplying solutions~\eqref{eq:AE_Sol1} and~\eqref{eq:AE_Sol2}, inserting 
the above mentioned correlators, and performing the integrals,  one obtains result~\eqref{eq:AE_equality2}.

\subsection{Computation of the right-hand side of Eq.~\eqref{eq:AE_equality}}
\label{subapp:AppendixRight}
For the right-hand side of Eq.~\eqref{eq:AE_equality}, we need the following correlation functions: 
$ {\langle x(t)y(t)x(t')y(t') \rangle}_{\rm st} $ and $ {\langle x(t)y(t)\cos\varphi(t')y(t') \rangle}_{\rm st} $, where $ t \geq t' $. For 
$ {\langle x(t)y(t)x(t')y(t') \rangle}_{\rm st} $, the relevant nonzero correlators are those given by Eq.~\eqref{eq:AE_coscos_sinsin_unsheared}, 
$ \langle f_x(s_1)f_x(s_2) \rangle = \langle f_y(s_1)f_y(s_2) \rangle = \frac{2D_{\rm t}}{\mu_{\rm t}^2}\delta(s_1-s_2) $, and
\begin{align}
\notag &{\langle \cos\varphi(s_1)\sin\varphi(s_2)\cos\varphi(s_3)\sin\varphi(s_4) \rangle}_{\rm st} = \frac{1}{8}\exp\Big\{-D_{\rm r}\big[s_1+s_2+s_3+s_4+2\min(s_1, s_2)-2\min(s_1, s_3)\\
\notag & - 2\min(s_1, s_4)-2\min(s_2, s_3)-2\min(s_2, s_4)+2\min(s_3, s_4)\big]\Big\}+ \frac{1}{8}\exp\Big\{-D_{\rm r}\big[s_1+s_2+s_3+s_4-2\min(s_1, s_2)\\
\notag &-2\min(s_1, s_3)+ 2\min(s_1, s_4)+2\min(s_2, s_3)-2\min(s_2, s_4)-2\min(s_3, s_4)\big]\Big\} - \frac{1}{8}\exp\Big\{-D_{\rm r}\big[s_1+s_2+s_3+s_4\\
& -2\min(s_1, s_2)+2\min(s_1, s_3)- 2\min(s_1, s_4)-2\min(s_2, s_3)+2\min(s_2, s_4)-2\min(s_3, s_4)\big]\Big\},
\label{eq:AE_cossincossin}
\end{align}
where $ \min(s_1, s_2) = s_1 $ if $ s_1 < s_2 $ and $ \min(s_1, s_2) = s_2 $ if $ s_2 < s_1 $. Note that the second term in Eq.~\eqref{eq:AE_cossincossin} equals minus the third one with either 
$ s_1 $ and $ s_2  $ or $ s_3 $ and $ s_4 $ interchanged. This leads to cancelation of these terms being integrated over either $ s_1 $ and $ s_2 $ or $ s_3 $ and $ s_4 $ in the same range. Therefore, 
these terms do not contribute to $ {\langle x(t)y(t)x(t')y(t') \rangle}_{\rm st} $ or to $ {\langle x(t)y(t)\cos\varphi(t')y(t') \rangle}_{\rm st} $. The final result for $ {\langle x(t)y(t)x(t')y(t') \rangle}_{\rm st} $ reads as
\begin{align}
\notag & {\langle x(t)y(t)x(t')y(t') \rangle}_{\rm st} = \frac{D_{\rm t}^2}{{(\mu_{\rm t} k)}^2}e^{-2\mu_{\rm t} k(t-t')} + \frac{v_0^2D_{\rm t}}{{(\mu_{\rm t} k)}^2\left[D_{\rm r}^2-{(\mu_{\rm t} k)}^2\right]}\left\{D_{\rm r}e^{-2\mu_{\rm t} k(t-t')}-\mu_{\rm t} ke^{-(D_{\rm r}+\mu_{\rm t} k)(t-t')}\right\}\\
\notag & + \frac{v_0^4}{8{(\mu_{\rm t} k)}^2\left[D_{\rm r}^2-{(\mu_{\rm t} k)}^2\right]\left[4D_{\rm r}^2-{(\mu_{\rm t} k)}^2\right]\left[3D_{\rm r}-\mu_{\rm t} k\right]\left[D_{\rm r}+3\mu_{\rm t} k\right]} \Bigg\{2D_{\rm r}^2\left[3D_{\rm r}-\mu_{\rm t} k\right]\left[4D_{\rm r} + 5\mu_{\rm t} k\right]e^{-2\mu_{\rm t} k(t-t')}\\
& \ \ \ \ - 12D_{\rm r}\mu_{\rm t} k\left[4D_{\rm r}^2-{(\mu_{\rm t} k)}^2\right]e^{-(D_{\rm r} + \mu_{\rm t} k)(t-t')} + {(\mu_{\rm t} k)}^2\left[D_{\rm r}-\mu_{\rm t} k\right]\left[D_{\rm r}+3\mu_{\rm t} k\right]e^{-4D_{\rm r}(t-t')}\Bigg\},
\label{eq:AE_xyxy}
\end{align}
where the first term is the result for a passive particle, the second term results from coupling between active motion and translational diffusion, and the third term is a purely active contribution.
For $ {\langle x(t)y(t)\cos\varphi(t')y(t') \rangle}_{\rm st} $, the relevant nonzero correlators are 
$ \langle f_x(s_1)f_x(s_2) \rangle = \langle f_y(s_1)f_y(s_2) \rangle = \frac{2D_{\rm t}}{\mu_{\rm t}^2}\delta(s_1-s_2) $ and those given by Eqs.~\eqref{eq:AE_coscos_sinsin_unsheared} and~\eqref{eq:AE_cossincossin}. We get 
\begin{align}
\notag & {\langle x(t)y(t)\cos\varphi(t')y(t') \rangle}_{\rm st} = \frac{v_0D_{\rm t}}{2\mu_{\rm t} k\left[D_{\rm r}^2-{(\mu_{\rm t} k)}^2\right]}\left\{2D_{\rm r}e^{-2\mu_{\rm t} k(t-t')}-\left(D_{\rm r}+\mu_{\rm t} k\right)e^{-(D_{\rm r}+\mu_{\rm t} k)(t-t')}\right\}\\
\notag & + \frac{v_0^3}{8\mu_{\rm t} k\left[D_{\rm r}^2-{(\mu_{\rm t} k)}^2\right]\left[4D_{\rm r}^2-{(\mu_{\rm t} k)}^2\right]\left[3D_{\rm r}-\mu_{\rm t} k\right]\left[D_{\rm r}+3\mu_{\rm t} k\right]} \Bigg\{4D_{\rm r}^2\left[3D_{\rm r}-\mu_{\rm t} k\right]\left[4D_{\rm r} + 5\mu_{\rm t} k\right]e^{-2\mu_{\rm t} k(t-t')}\\
& \ \ \ \ - 6D_{\rm r}\left[4D_{\rm r}^2-{(\mu_{\rm t} k)}^2\right]\left[D_{\rm r} + 3\mu_{\rm t} k\right]e^{-(D_{\rm r} + \mu_{\rm t} k)(t-t')} + \mu_{\rm t} k\left[D_{\rm r}-\mu_{\rm t} k\right]\left[2D_{\rm r}+\mu_{\rm t} k\right]\left[D_{\rm r}+3\mu_{\rm t} k\right]e^{-4D_{\rm r}(t-t')}\Bigg\}.
\label{eq:AE_xycosy}
\end{align}

Using Eqs.~\eqref{eq:AE_xyxy} and~\eqref{eq:AE_xycosy}, we find for the three terms on the right-hand side of Eq.~\eqref{eq:AE_equality}
\begin{align}
\notag & \frac{\dot{\gamma}}{4D_{\rm t}}{\left\langle x(t)y(t)\left[x(t)y(t)-x(0)y(0)\right]\right\rangle}_{\rm st} = \frac{\dot{\gamma}D_{\rm t}}{4{(\mu_{\rm t} k)}^2}\left(1-e^{-2\mu_{\rm t} kt}\right) + \frac{\dot{\gamma}v_0^2}{4{(\mu_{\rm t} k)}^2\left[D_{\rm r}^2-{(\mu_{\rm t} k)}^2\right]}\Bigg\{D_{\rm r}\left(1-e^{-2\mu_{\rm t} kt}\right)\\
\notag & - \mu_{\rm t} k\left(1-e^{-(D_{\rm r}+\mu_{\rm t} k)t}\right)\Bigg\} + \frac{\dot{\gamma}v_0^4}{32D_{\rm t}{(\mu_{\rm t} k)}^2\left[D_{\rm r}^2-{(\mu_{\rm t} k)}^2\right]\left[4D_{\rm r}^2-{(\mu_{\rm t} k)}^2\right]\left[3D_{\rm r}-\mu_{\rm t} k\right]\left[D_{\rm r}+3\mu_{\rm t} k\right]}\\
\notag & \ \ \ \ \times \Bigg\{2D_{\rm r}^2\left[3D_{\rm r}-\mu_{\rm t} k\right]\left[4D_{\rm r} + 5\mu_{\rm t} k\right]\left(1-e^{-2\mu_{\rm t} kt}\right) - 12D_{\rm r}\mu_{\rm t} k\left[4D_{\rm r}^2-{(\mu_{\rm t} k)}^2\right]\left(1-e^{-(D_{\rm r} + \mu_{\rm t} k)t}\right)\\
& \ \ \ \ \ \ \ \ + {(\mu_{\rm t} k)}^2\left[D_{\rm r}-\mu_{\rm t} k\right]\left[D_{\rm r}+3\mu_{\rm t} k\right]\left(1-e^{-4D_{\rm r}t}\right)\Bigg\},\label{eq:AE_equality3_1}\\
\notag & - \frac{\dot{\gamma}v_0}{2D_{\rm t}}\int_0^tdt'{\left\langle x(t)y(t)\cos\varphi(t')y(t')\right\rangle}_{\rm st} = - \frac{\dot{\gamma}v_0^2}{4{(\mu_{\rm t} k)}^2\left[D_{\rm r}^2-{(\mu_{\rm t} k)}^2\right]}\Bigg\{D_{\rm r}\left(1-e^{-2\mu_{\rm t} kt}\right)- \mu_{\rm t} k\left(1-e^{-(D_{\rm r}+\mu_{\rm t} k)t}\right)\Bigg\}\\
\notag & - \frac{\dot{\gamma}v_0^4}{16D_{\rm t}{(\mu_{\rm t} k)}^2\left[D_{\rm r}^2-{(\mu_{\rm t} k)}^2\right]\left[4D_{\rm r}^2-{(\mu_{\rm t} k)}^2\right]\left[3D_{\rm r}-\mu_{\rm t} k\right]\left[D_{\rm r}+3\mu_{\rm t} k\right]}\\
\notag & \ \ \ \ \times \Bigg\{2D_{\rm r}^2\left[3D_{\rm r}-\mu_{\rm t} k\right]\left[4D_{\rm r} + 5\mu_{\rm t} k\right]\left(1-e^{-2\mu_{\rm t} kt}\right) - \frac{6D_{\rm r}\mu_{\rm t} k}{D_{\rm r}+\mu_{\rm t} k}\left[4D_{\rm r}^2-{(\mu_{\rm t} k)}^2\right]\left[D_{\rm r}+3\mu_{\rm t} k\right]\left(1-e^{-(D_{\rm r} + \mu_{\rm t} k)t}\right)\\
& \ \ \ \ \ \ \ \ + \frac{{(\mu_{\rm t} k)}^2}{4D_{\rm r}}\left[D_{\rm r}-\mu_{\rm t} k\right]\left[2D_{\rm r}+\mu_{\rm t} k\right]\left[D_{\rm r}+3\mu_{\rm t} k\right]\left(1-e^{-4D_{\rm r}t}\right)\Bigg\},\label{eq:AE_equality3_2}\\
\notag & \frac{\dot{\gamma}\mu_{\rm t} k}{2D_{\rm t}}\int_0^tdt'{\left\langle x(t)y(t)x(t')y(t')\right\rangle}_{\rm st} = \frac{\dot{\gamma}D_{\rm t}}{4{(\mu_{\rm t} k)}^2}\left(1 - e^{-2 \mu_{\rm t} k t}\right) + \frac{\dot{\gamma}v_0^2}{4{(\mu_{\rm t} k)}^2\left[D_{\rm r}^2-{(\mu_{\rm t} k)}^2\right]}\Bigg\{D_{\rm r}\left(1-e^{-2\mu_{\rm t} kt}\right)\\
\notag & - \frac{2{(\mu_{\rm t} k)}^2}{D_{\rm r}+\mu_{\rm t} k}\left(1-e^{-(D_{\rm r}+\mu_{\rm t} k)t}\right)\Bigg\} + \frac{\dot{\gamma}v_0^4}{16D_{\rm t}{(\mu_{\rm t} k)^2}\left[D_{\rm r}^2-{(\mu_{\rm t} k)}^2\right]\left[4D_{\rm r}^2-{(\mu_{\rm t} k)}^2\right]\left[3D_{\rm r}-\mu_{\rm t} k\right]\left[D_{\rm r}+3\mu_{\rm t} k\right]}\\
\notag & \times \Bigg\{D_{\rm r}^2\left[3D_{\rm r}-\mu_{\rm t} k\right]\left[4D_{\rm r} + 5\mu_{\rm t} k\right]\left(1-e^{-2\mu_{\rm t} kt}\right)- \frac{12D_{\rm r}{(\mu_{\rm t} k)}^2}{D_{\rm r}+\mu_{\rm t} k}\left[4D_{\rm r}^2-{(\mu_{\rm t} k)}^2\right]\left(1-e^{-(D_{\rm r} + \mu_{\rm t} k)t}\right)\\
& \ \ \ \ + \frac{{(\mu_{\rm t} k)}^3}{4D_{\rm r}}\left[D_{\rm r}-\mu_{\rm t} k\right]\left[D_{\rm r}+3\mu_{\rm t} k\right]\left(1-e^{-4D_{\rm r}t}\right)\Bigg\}.\label{eq:AE_equality3_3}
\end{align}
Adding the right-hand sides of Eqs.~\eqref{eq:AE_equality3_1} --~\eqref{eq:AE_equality3_3} together, one finds that the terms proportional to 
$ \frac{\dot{\gamma}v_0^4}{D_{\rm t}{(\mu_{\rm t} k)}^2} $ cancel and the rest gives Eq.~\eqref{eq:AE_equality2}. This completes our check of Eq.~\eqref{eq:AE_equality}.

In addition, we also checked that $ {\langle x(t)y(t)\cos\varphi(t')y(t') \rangle}_{\rm st} = {\langle x(t)y(t)\sin\varphi(t')x(t') \rangle}_{\rm st} $, thereby confirming, for this specific example, our statement in Subsec.~\ref{subsec:LR_another_form} regarding the fact that the terms $\cos\varphi_i(t')y_i(t')$ and $\sin\varphi_i(t')x_i(t')$ in Eq.~\eqref{eq:LRAF2} give identical contributions.
\twocolumngrid

\end{document}